\documentclass[12pt]{article}
\pdfoutput=1

\usepackage{amssymb}
\usepackage{color}
\usepackage{graphicx,psfrag}
\usepackage{amsmath}
\usepackage{hyperref}

\newcommand{\refeq}[1]{(\ref{#1})}

\newcommand{\R}{\mathcal{R}}

\newcommand{\ka}{\mathbf{k}}

\def\Tdot#1{{{#1}^{\hbox{.}}}}

\def\s{s_1}

\def\ps{\psi}

\usepackage{soul}
\definecolor{lm}{cmyk}{.20, 0, .30, 0}
\sethlcolor{lm}

\def\be{\begin{equation}}
\def\ee{\end{equation}}
\def\bea{\begin{eqnarray}}
\def\eea{\end{eqnarray}}

\begin{document}

\begin{titlepage}

\setcounter{page}{1} \baselineskip=15.5pt \thispagestyle{empty}

\bigskip\
\begin{center}
{\Large \bf  A Statistical Approach to Multifield Inflation:}
\vskip 8pt
{\Large \bf  Many-field Perturbations Beyond Slow Roll}
\vskip 5pt
\vskip 15pt
\end{center}
\vspace{0.5cm}
\begin{center}
{
Liam McAllister,${}^1$  S\'ebastien Renaux-Petel,${}^{2,3}$ and Gang Xu${}^{1,4}$}
\end{center}

\vspace{0.1cm}

\begin{center}
\vskip 4pt
\textsl{$^{1}$  Department of Physics, Cornell University,
Ithaca, NY 14853 USA}
\vskip 4pt
\textsl{$^{2}$  Centre for Theoretical Cosmology,\\
Department of Applied Mathematics and Theoretical Physics,\\
University of Cambridge, Cambridge CB3 0WA, UK}
\vskip 4pt
\textsl{$^{3}$  UPMC Univ Paris 06, CNRS, Institut Lagrange de Paris,\\
Laboratoire de Physique Th\'eorique et Hautes Energies,\\
UMR 7589, 4 place Jussieu, 75252 Paris Cedex 05, France}
\vskip 4pt
\textsl{$^{4}$  Perimeter Institute for Theoretical Physics,\\
31 Caroline Street North, Waterloo ON, Canada}

\end{center} 
{\small  \noindent  \\[0.2cm]
\noindent

We study multifield contributions to the scalar power spectrum in an ensemble of six-field inflationary models obtained in string theory.
We identify examples in which inflation occurs by chance, near an approximate inflection point, and
we compute the primordial perturbations numerically, both exactly and using an array of truncated models.
The scalar mass spectrum and the number of fluctuating fields are accurately described by a simple random matrix model.
During the approach to the inflection point, bending trajectories and violations of slow roll are commonplace,
and `many-field' effects, in which three or more fields influence the perturbations, are often important.
However, in a large fraction of models consistent with constraints on the tilt the
signatures of multifield evolution occur on unobservably large scales. Our scenario is a concrete microphysical realization of quasi-single-field inflation, with scalar masses of order $H$, but the cubic and quartic couplings are typically too small to produce detectable non-Gaussianity.
We argue that our results are characteristic of a broader class of models arising from  multifield potentials that are natural in the Wilsonian sense.}

\vspace{0.3cm}

\vspace{0.6cm}

\vfil
\begin{flushleft}
\small \today
\end{flushleft}
\end{titlepage}

\newpage
\tableofcontents
\newpage

\section{Introduction}

Inflation \cite{Guth:1980zm,Linde:1981mu,Albrecht:1982wi} provides a superb explanation for the observed spectrum of cosmic microwave background (CMB) anisotropies.
The simplest and best-understood models of inflation involve a single field slowly rolling down a relatively flat potential, but more complicated models involving multiple light fields and/or violations of slow roll are arguably more natural, and provide the prospect of distinctive signatures.

Strong theoretical arguments motivate the consideration of inflationary models with many light fields. In theories with spontaneously broken supersymmetry, naturalness suggests that scalars receive masses that are at least of order the gravitino mass.  If the inflationary energy is the dominant source of supersymmetry breaking in the early universe,
scalars
that couple with gravitational strength to this energy will acquire
masses of order the inflationary Hubble parameter, $H$.  Thus, moduli in the theory do not decouple from the inflationary dynamics, and can be light enough to fluctuate.\footnote{See \cite{BaumannGreen} for a discussion of some of the cosmological signatures of models with spontaneously broken supersymmetry.}  This picture is well-attested in flux compactifications of string theory, which typically include tens or hundreds of moduli with masses clustered around $H$.
In cases where the moduli potential is computable, one generally finds a complicated, high-dimensional potential energy landscape with structure dictated by the spectrum of Planck-suppressed operators in the theory.
The nature of inflation in such a potential is an important and urgent problem.

However, despite intensive efforts to understand multifield inflation over the past decade, most analyses of explicit models consider only two light fields.  Moreover, even though significant analytical tools have been developed to trace the evolution of primordial perturbations outside the Hubble radius in models violating the slow roll approximation (see \S\ref{decomposition} for a brief review of prior results), a large majority of works on the subject do make a slow roll expansion during Hubble exit.  An understanding of truly general models involving several fields with masses of order $H$
and arbitrary dynamics remains necessary.

In this work we study the primordial perturbations produced by inflation in a class of six-field potentials obtained in string theory \cite{Baumann10}, corresponding to a D3-brane moving in a conifold region of a stabilized compactification.\footnote{This system was recently studied by Dias, Frazer, and Liddle in \cite{Dias}.  In \S\ref{isocurvature} we explain how our conclusions are qualitatively different from those of \cite{Dias}: in contrast to \cite{Dias}, we find that an adiabatic limit is reached during inflation, so that the curvature perturbations can be predicted without modeling the details of reheating.}
Our approach is statistical: instead of directly fine-tuning the potential to achieve inflation, we draw potentials at random from a well-specified ensemble and study realizations that inflate by chance.
We then compute the exact dynamics of the linear perturbations numerically, making no slow roll approximation.  To reveal the physical processes underlying the resulting perturbations, we recompute the perturbations using a range of approximate, truncated descriptions that retain different subsets of the entropic modes, and we then cross-correlate the exact and approximate answers.  For example, we identify multiple-field contributions to the perturbations by comparing the exact spectrum to the spectrum calculated in a single-field truncation (in which no slow roll approximation is made).

The organization of this paper is as follows.  In \S\ref{sec:method} we describe our method for generating inflationary trajectories and computing the corresponding primordial perturbations.
In \S\ref{sec:spectrum} we show that the scalar mass spectrum follows from a simple matrix model, and we assess the incidence and consequences of slow roll violations.
In \S\ref{multifield} we study two-field and many-field contributions to the scalar power spectrum.
We conclude in \S\ref{conclusions}.

\section{Method} \label{sec:method}

In this section we briefly review warped D-brane inflation, describe how we identify an ensemble of inflating solutions, and then explain how we compute the primordial perturbations.

\subsection{Background evolution in D-brane inflation}

In warped D-brane inflation \cite{KKLMMT}, inflation is driven by a D3-brane moving toward an anti-D3-brane in a warped throat region of a flux compactification.  The structure of the potential for this configuration has been derived in \cite{Baumann08, Baumann10}.
The operators in the effective Lagrangian are dictated by the throat geometry, and have been computed explicitly, but the Wilson coefficients are determined by the detailed configuration of sources (fluxes, Euclidean D-branes, etc.) in the bulk of the compactification.  Thus, with present knowledge the Wilson coefficients can at best be modeled statistically.

In \cite{Agarwal:2011wm}, two of us, in collaboration with N.~Agarwal and R.~Bean, studied a large number of realizations of the potential, drawing the Wilson coefficients from statistical distributions, and found that although the typical scale of all six scalar masses is $H$, accidental cancellations among many terms could nevertheless lead to prolonged inflation.  The inflationary trajectory took a characteristic form: the D3-brane initially moved rapidly in the angular directions of the conifold, spiraled down to an inflection point in the potential, and then settled into an inflating phase.  It was established in  \cite{Agarwal:2011wm} that the inflationary phenomenology has negligible dependence on the detailed form of the statistical distribution of the Wilson coefficients: a sort of universality emerges in this complicated ensemble.\footnote{See \cite{Panda:2007ie,Easson:2007dh,Langlois:2008wt,Langlois:2008qf,Chen:2008ada,Gao:2009gd,Langlois:2009ej,Mizuno:2009cv,RenauxPetel:2009sj,Mizuno:2009mv,Chen:2010qz,Gregory:2011cd,RenauxPetel:2011uk,Dias} for related work on multifield evolution in D-brane inflation.}

Our method for studying the evolution of the homogeneous background is identical to that of \cite{Agarwal:2011wm}, to which we refer for further details.
We began with an ensemble of more than 13 million potentials describing a D3-brane in a conifold geometry.
For each potential we started with zero kinetic energy at a fixed location in the conifold\footnote{Trajectories with nonvanishing initial  kinetic energy were studied in \cite{Agarwal:2011wm}, where it was found that as long as the initial kinetic energy is no larger than the initial potential energy, the effect on the phenomenology is minimal. Similarly, it was shown in \cite{Agarwal:2011wm} that the dependence on the initial radial location is minimal, while the rotational invariance of our ensemble of potentials implies that we do not lose any generality by fixing the initial angular positions.} and evolved the background equations of motion numerically, as detailed in \cite{Agarwal:2011wm}.  After discarding trials in which the inflaton became stuck in a local minimum or was ejected from the throat region, we obtained 18731 realizations yielding at least 66 e-folds\footnote{Although we are interested in requiring at least 60 e-folds of expansion to solve the horizon problem, we insist on a total of 66 e-folds in order to be able to impose Bunch-Davies initial conditions well before the CMB exits the Hubble radius.} of expansion
followed by a hybrid exit.\footnote{In practice we take the end of inflation to occur when the D3-brane reaches a fixed radial location slightly above the tip.}

A word of caution about choices of measure is necessary.  As shown in \cite{Agarwal:2011wm}, the inflationary phenomenology is substantially independent of the statistical distribution from which the Wilson coefficients are drawn, and of the measure taken on the space of homogeneous initial conditions.
However, we do not include
a measure factor that weights histories according to the amount of expansion, and this must be borne in mind when interpreting our results.

\subsection{Aspects of inflection point inflation} \label{inflection}

When inflation arises in the ensemble of potentials considered in this work, it does so at an approximate inflection point \cite{delicate,Krause:2007jk, Baumann:2007ah, Agarwal:2011wm}.  Before proceeding we will recall a few  elementary properties of single-field inflection point inflation that have significant ramifications for our analysis.

\begin{figure}[!h]
  \begin{center}
    \includegraphics[width=4.0in,angle=0]{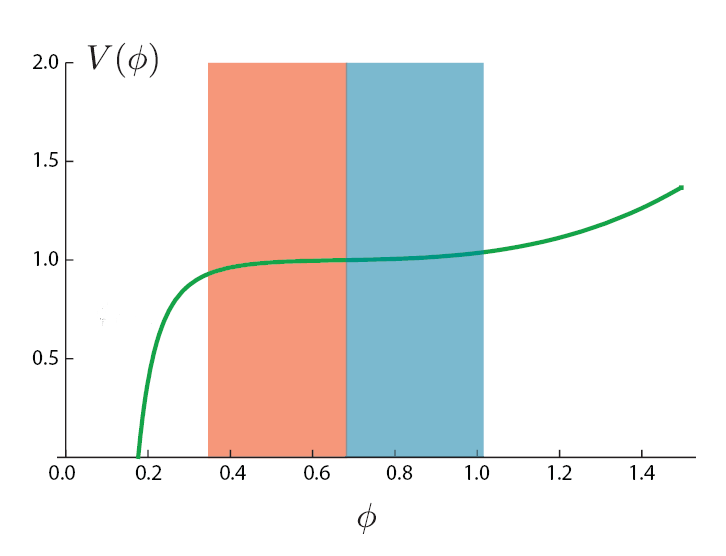}
    \caption{A single-field inflection point potential, taken from \cite{Baumann:2007ah}.  If the inflaton is above the inflection point 60 e-folds before the end of inflation, $V''> 0$ and the scalar power spectrum is blue, as indicated by the shading.  A red spectrum requires that the inflaton has passed the inflection point 60 or more e-folds before the end of inflation.}
    \label{fig:inflection}
  \end{center}
\end{figure}

Inflection point inflation begins in a region of field space where the potential is positively curved and evolves to a region where the curvature is negative (see Fig.~\ref{fig:inflection}). As the size of the inflection point region in Planck units is typically extremely small (see e.g.\ \cite{Agarwal:2011wm}), the potential slow roll parameters $\epsilon_V \equiv \frac{1}{2}M_{p}^2 \left(\frac{V'}{V}\right)^2$ and  $\eta_V \equiv M_{p}^2 \frac{V''}{V}$ obey $\epsilon_V \ll | \eta_V |$, so that $n_{s} \approx 1 + 2\eta_V$.  Hence, the scalar power spectrum calculated in the single-field slow roll approximation is initially blue and becomes red.  Correspondingly, the tilt measured in the CMB is dictated by the point in field space at which observable modes exit the Hubble radius.  Inflection points producing a total of $N_e \approx 120$ e-folds of inflation have approximately 60 e-folds below the inflection point and 60 e-folds above, so that the primordial spectrum on large angular scales is scale-invariant. Inflection points producing $N_e > 120$ e-folds have red spectra, while inflection points producing $N_e < 120$ e-folds have blue spectra and are in tension with  observations.

\subsection{Perturbations in multifield inflation} \label{reviewmulti}
In this section, we review an efficient approach to computing the primordial perturbations in multifield inflation, following \cite{GrootNibbelink:2001qt} and \cite{Langlois:2008mn}.

\subsubsection{Equations of motion for exact multifield treatment }\label{multifieldQs}

Our starting point  is the action  for a collection of scalar fields $\phi^I$, endowed with a metric\footnote{Relativistic motion, corresponding to Dirac-Born-Infeld inflation \cite{Silverstein:2003hf,Alishahiha:2004eh}, did not arise in our ensemble \cite{Bean:2007hc,Agarwal:2011wm}, so it suffices to consider the two-derivative kinetic term.}
$G_{IJ}(\phi^K)$ on field space,
interacting through 	a potential $V(\phi^I)$, and minimally coupled to gravity:
\begin{eqnarray}
S= \int  d^4x \,\sqrt{-g} \left( \frac{R}{2}  -\frac{1}{2}G_{IJ } \nabla_{\mu} \phi^I \nabla^{\mu} \phi^J -V(\phi^I) \right)
\label{S2}
\end{eqnarray}
(see \cite{Candelas:1989js} for the explicit expression for the field space metric in the case of the conifold).

The background metric is assumed to be of the spatially flat Friedmann-Lema\^itre-Robertson-Walker form
\begin{equation}
ds^2=-dt^2+a^2(t)d{\vec x}^2\,,
\end{equation}
where $t$ is cosmic time and $a(t)$ denotes the scale factor.  If the fields  depend only on  $t$, the equations of motion take the simple form
\begin{eqnarray}
 3H^2&=&\frac12 \dot \sigma^2+V\,,  \\
 \dot{H}&=&-\frac12\dot \sigma^2\,, \\
{\cal D}_t \dot \phi^I  +3H  \dot \phi^I+G^{IJ} V_{,J}&=&0\,,
\end{eqnarray}
where dots denote derivatives with respect to $t$, $H \equiv \dot a/a$ is the Hubble parameter, $\frac{1}{2}\dot \sigma^2 \equiv \frac{1}{2}G_{IJ} \dot \phi^I \dot \phi^J$ is the kinetic energy of the fields, and, here and in the following, ${\cal D}_t A^I \equiv \dot{A^I} + \Gamma^I_{JK} \dot \phi^J A^K$ for a field space vector $A^I$.

The dynamics of linear
perturbations about this background is governed by the second-order action \cite{Sasaki:1995aw,GrootNibbelink:2001qt,Langlois:2008mn}
 \begin{eqnarray}
S_{(2)}= \int  dt\, d^3x \,a^3\left(G_{IJ}\mathcal{D}_tQ^I\mathcal{D}_tQ^J-\frac{1}{a^2}G_{IJ}\partial_i Q^I \partial^i Q^J-M_{IJ}Q^IQ^J\right)\,,
\label{S2}
\end{eqnarray}
where the $Q^I$ are the field fluctuations in the spatially flat gauge and the mass (squared) matrix is given by
\begin{eqnarray} \label{masssquared}
M_{IJ} &=& V_{; IJ} - \mathcal{R}_{IKLJ}\dot \phi^K \dot \phi^L -\frac{1}{a^3}\mathcal{D}_t\left[\frac{a^3}{H} \dot \phi_
I \dot \phi_J\right]\,.
\end{eqnarray}
Here $ V_{;IJ} \equiv V_{,IJ}-\Gamma_{IJ}^K V_{,K}$ is the covariant Hessian, $\mathcal{R}_{IKLJ}$ is the Riemann tensor associated to the field space metric, and field space indices are raised and lowered using
$G_{IJ}$.
From equation \refeq{S2} one easily deduces the equations of motion for the linear fluctuations (in Fourier space):
\begin{eqnarray}
{\cal D}_t {\cal D}_t Q^I  +3H {\cal D}_t Q^I +\frac{k^2}{a^2} Q^I +M^I_J Q^J=0\,.
\label{pert}
\end{eqnarray}

\subsubsection{The adiabatic/entropic decomposition} \label{decomposition}

Following \cite{Gordon:2000hv,GrootNibbelink:2001qt}, it is useful to decompose
the field fluctuations into the so-called (instantaneous) adiabatic and entropic perturbations. The adiabatic perturbation is defined as
\begin{equation}
Q_{\sigma} \equiv e_{\sigma I} Q^I\,,
\end{equation}
where $e_{\sigma}^I \equiv \dot \phi^I /{\dot \sigma}$ is the unit vector pointing along the background trajectory in field space, whereas entropic fluctuations represent fluctuations off the background trajectory. The adiabatic fluctuation
is directly proportional to the comoving curvature perturbation $\cal R$,
\begin{eqnarray}
{\cal R} =\frac{H}{{\dot \sigma}} Q_{\sigma}\,,
\end{eqnarray}
while the genuinely multifield effects are embodied by the entropic fluctuations.

One of the entropic modes plays a distinguished role: this is the `first' entropic fluctuation $Q_{\s} \equiv e_{\s I} Q^I$,
which is the fluctuation
along the direction of acceleration perpendicular to the background trajectory, where
\begin{equation}
e_{\s}^I \equiv  -\frac{\perp^{IJ}V_{,J} }{ \sqrt{\perp^{IJ} V_{,I} V_{,J}}}\,,
\end{equation}
and $\perp^{IJ} \equiv G^{IJ}-e_{\sigma}^{I} e_{\sigma}^J$ is the projection operator on the entropic subspace.
The first entropic mode instantaneously couples to the adiabatic perturbation, but the remaining entropic modes do not.

The adiabatic equation of motion can be written in the compact form
\begin{eqnarray}
\label{Qsigma}
 \ddot{Q}_{\sigma}+3H
 \dot{Q}_{\sigma}+\left(\frac{k^2}{a^2}+m_{\sigma}^2\right)  Q_{\sigma} = \Tdot{\left(2 H \eta_\perp Q_{\s}\right)}
-\left(\frac{\dot H}{H}
+\frac{V_{,\sigma}}{\dot \sigma }\right)  2 H \eta_\perp\, Q_{\s}\,,
\end{eqnarray}
where
\begin{equation} \label{etaperpdefinition}
\eta_\perp  \equiv -\frac{V_{,\s} }{H \dot \sigma}
\end{equation}
is a very important dimensionless parameter measuring the size of the coupling between the adiabatic mode and the first entropic mode.
Here $V_{,\s}  \equiv e_{\s}^I V_{,I}$, and similarly for analogous quantities, and the adiabatic mass (squared)
$m_{\sigma}^2$ is given by
\bea
\frac{m_\sigma^2}{H^2} \equiv 3\eta-\eta^2+\epsilon \eta +\dot \eta/H\,,
\eea
with $\epsilon \equiv -\frac{\dot H}{H^2}$ and $\eta \equiv -\frac12\frac{\dot \epsilon}{H \epsilon}$.

A brief overview of existing methods to study cosmological fluctuations in multifield inflation is appropriate at this stage.  Several analytical methods have been developed to follow the evolution of perturbations outside the Hubble radius without making any slow roll approximation, including the $\delta N$ formalism \cite{Starobinsky:1986fxa,Sasaki:1995aw,Sasaki:1998ug,Lyth:2005fi},
the transfer functions method \cite{transfer}, the gradient expansion method \cite{Rigopoulos:2004ba,Rigopoulos:2005xx,Rigopoulos:2005ae,Rigopoulos:2005us}, the covariant method \cite{Langlois:2006vv,RenauxPetel:2008gi,Lehners:2009ja}, and the transport equations method \cite{Mulryne:2009kh,Mulryne:2010rp,Dias:2011xy,Anderson:2012em}.  A number of works incorporate perturbative corrections to the slow roll approximation  during the epoch of Hubble exit, including \cite{GrootNibbelink:2001qt,Lee:2005bb,Byrnes:2006fr,Lalak:2007vi,Choi:2007su,Avgoustidis:2011em}, but exact results are scarce for systems with more than two fluctuating fields.
Related investigations of the effects of heavy fields in multifield inflation include \cite{Cremonini:2010ua,Achucarro:2010da,Baumann:2011su,Shiu:2011qw,Cespedes:2012hu,Avgoustidis:2012yc, Achucarro:2012yr, Gao:2012uq}, while numerical studies for two-field systems include \cite{Tsujikawa:2002qx,Lalak:2007vi,Peterson:2010np}.

We stress that an exact numerical treatment is needed, at present, for models involving significant bending around Hubble crossing, which as we will see are common in our ensemble.

\subsubsection{Methods for computing the perturbations} \label{pertmethod}

For each of the 18731 realizations of inflation yielding at least 66 e-folds of expansion followed by a hybrid exit, we numerically integrated the exact equations of motion (\ref{pert}) for the linearized perturbations corresponding to the scale exiting the Hubble radius 60 e-folds before the end of inflation.  To relate this result to conceptually
simpler models, we also computed the perturbations using an array of approximate descriptions, which we now specify.  To distinguish references to these specific models from more general uses of words such as ``exact'', ``naive'', etc., we will  put the {\sf model name}
in a distinctive font.

The {\sf naive} model simply computes $\frac{H^2}{8\pi^2\epsilon}$ at Hubble exit,\footnote{All quantities described as `evaluated at  Hubble exit' in this work are actually averaged from one e-fold before Hubble exit until one e-fold after Hubble exit, in order to smooth quantities that may be rapidly changing.} and incorporates neither slow roll violations nor multifield effects.
The adiabatic or {\sf{one-field}}  model makes no slow roll approximation, but discards the effects of all entropic modes: only fluctuations tangent to the trajectory at any given time are retained.

The {\sf{two-field}} model keeps the instantaneous
adiabatic and first entropic fluctuations.  For this case  the first entropic equation of motion takes the form
\begin{eqnarray}
 \ddot{Q}_{\s}+ 3H \dot{Q}_{\s}+\left(\frac{k^2}{a^2}+m_{\s}^2\right)Q_{\s}=-2 \dot \sigma \eta_\perp \dot {\cal R}\,,
\label{delta_s_all_scales}
\end{eqnarray}
where
\begin{equation}  \label{delta_s_all_scales2}
m_{\s}^2\equiv V_{;\s \s}- \dot \sigma^2 R_{\s \sigma \sigma \s}- H^2 \eta_\perp^2 \,.
\end{equation}
The {\sf{three-field}}, {\sf{four-field}}, and {\sf{five-field}} models (which we will not use in this work)
retain additional entropic perturbations along corresponding basis vectors in the decomposition of \cite{GrootNibbelink:2001qt}.
We will refer to the {\sf{six-field}} model, which includes all five entropic modes and therefore incorporates all the physics of the linear perturbations, as the {\sf{exact}} model.
The scalar power spectra resulting in each model are named in a similar manner, e.g.\ ${\cal P}_{{\sf one}}$ is the scalar power computed in the  {\sf{one-field}}  model.

By comparing the results of these approximate descriptions, one can unambiguously identify certain physical effects in the perturbations.  Specifically,  ${\cal P}_{{\sf one}} \neq {\cal P}_{{\sf naive}}$
signifies violations of slow roll, while  ${\cal P}_{{\sf exact}} \neq {\cal P}_{{\sf k}}$
demonstrates that at least $k+1$ fields contributed to the perturbations.

Finally, a discussion about numerical implementation of the quantization and evolution of the perturbations is in order. As is well known (see for instance \cite{vanTent:2002az,Weinberg:2008zzc,Tsujikawa:2002qx}), to determine the late-time power spectrum by numerically evolving the coupled equations \refeq{pert},
one cannot simply impose a single choice of initial conditions and solve the system \refeq{pert} once to obtain the six perturbations $Q^I$. This would introduce spurious interference terms in the power spectrum between what are supposed to be independent variables. On the contrary, one should identify six variables that are independent deep inside the Hubble radius, each corresponding to an independent set of creation and annihilation operators, and solve the system of equations \refeq{pert} six times,\footnote{Equivalently, one can solve a matrix-valued system once --- see \cite{Salopek:1988qh, vanTent:2002az, Huston:2011fr}.
We thank Ian Huston for pointing out Ref.~\cite{Salopek:1988qh}.}  
each time imposing the Bunch-Davies initial conditions for only one of the independent variables, while setting the other variables to zero initially.\footnote{In the {\sf k-field} models with $k < 6$ described above, we set $6-k$ entropic modes to zero throughout the evolution, not just initially.}  One then extracts power spectra by summing the relevant quantities over all six runs. This is the numerical analogue of the fact that the various creation and annihilation operators are independent, so that their effects add incoherently (see \S\ref{destructive} for a precise illustration of this method).

Deep inside the Hubble radius, one can neglect the mass matrix in equation \refeq{S2}, so that identifying a set of independent variables is equivalent to identifying a set of vielbeins for the field space metric $G_{IJ}$, which is easily accomplished numerically.
We follow this strategy for the quantization, imposing initial conditions six
e-folds before Hubble crossing,
while still solving the system of equations \refeq{pert} in the natural coordinate basis on the conifold (i.e. $r,\theta_1,\phi_1,\theta_2,\phi_2,\psi$ --- see \cite{Candelas:1989js}), which we find numerically efficient.

The method above applies to the {\sf exact} model, but some modifications are required
for the {\sf k-field} models with $k < 6$, where by definition only $k$ independent variables are included.
While there is no possible subtlety for the {\sf one-field} model, for the {\sf two-field} model for instance, the easiest set of vielbeins
consists of the adiabatic and first entropic basis vectors. From a numerical perspective, the kinematical basis of \cite{GrootNibbelink:2001qt} corresponds to a particular set of vielbeins with transparent physical meaning.

\subsubsection{Terminology for restricted ensembles} \label{namesofensembles}

For convenience, we now define important subsets of our ensemble.
The full ensemble consists of all realizations yielding at least 66 e-folds of inflation.  The effectively single-field subset consists of all realizations in which multifield contributions to the scalar power are at most 1\% corrections: specifically, we require that
\begin{equation}
\xi_{{\sf exact}/{\sf one}} \equiv  |{\cal P}_{{\sf exact}}/{\cal P}_{{\sf one}}-1| < 0.01 \,.
\end{equation}
The effectively multifield ensemble is the complementary subset, consisting of all cases with $\xi_{{\sf exact}/{\sf one}}  \ge 0.01$.
Similarly, we define
\begin{equation}
\xi_{{\sf exact}/{\sf two}} \equiv  |{\cal P}_{{\sf exact}}/{\cal P}_{{\sf two}}-1|\,,
\end{equation} which measures the extent to which a two-field description is insufficient. The set of models with $\xi_{{\sf exact}/{\sf two}}  \ge 0.01$ may be termed `effectively many-field'.
Finally, the observationally allowed ensemble consists of all realizations satisfying the WMAP7 constraints on the tilt\footnote{See \S\ref{constraints} for a discussion of constraints on the amplitude and running of the scalar power spectrum.} of the scalar power spectrum at 2$\sigma$.

\section{The scalar mass spectrum and violations of slow roll}
\label{sec:spectrum}

Before describing the primordial perturbations, we will first characterize the spectrum of scalar masses.  We begin with empirical observations about the mass spectrum (\S\ref{spectrumobservations}), and then turn to obtaining several of these properties from a random matrix model (\S\ref{rmt}).  In \S\ref{srv} we present key consequences of the scalar mass spectrum, focusing on violations of slow roll and the resulting imprint in the power spectrum.

\subsection{Properties of the mass spectrum} \label{spectrumobservations}

\begin{figure}[!h]
  \begin{center}
    \includegraphics[width=5.0in,angle=0]{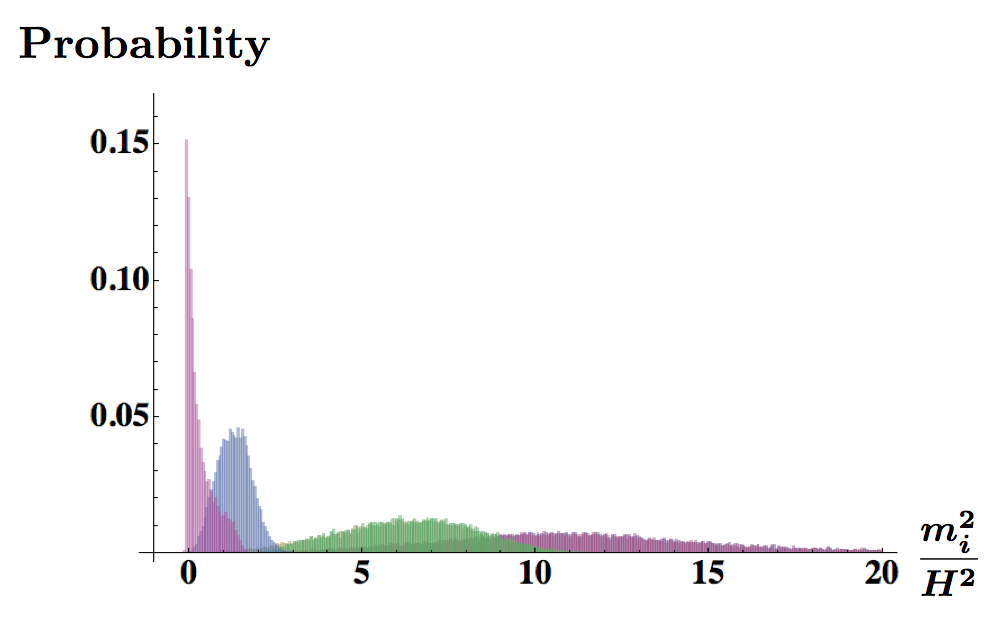}
    \caption{The mass spectra of the six scalar fields, in units of $H^2$.  The leftmost peak, which has support at tachyonic values, corresponds to the lightest (adiabatic) field $\ps_1$.  The next peak corresponds to the second-lightest field $\ps_2$.  The third peak corresponds to $\ps_3$ and $\ps_4$, which are nearly degenerate in each realization, and the broad final peak similarly corresponds to $\ps_5$ and $\ps_6$.}
    \label{spectrumofall}
  \end{center}
\end{figure}

For each inflationary trajectory we computed the Hessian matrix
$V^{;I}_{\,\,\,\,J}$
of the scalar potential, in terms of the canonical coordinates constructed at the relevant location on the conifold, at the moment that the CMB exited the Hubble radius. We denote the ordered eigenvalues of the Hessian as $m_1^2\le\ldots \le m_6^2$, with corresponding eigenvectors\footnote{In general the $\ps_i$ are complicated combinations of the six natural coordinates on the conifold, though in trajectories that are primarily radial there is significant overlap between $\ps_1$ and the radial coordinate $r$.} $\ps_1,\ldots \ps_6$.    A histogram of $m_1^2 \le \ldots \le m_6^2$ appears in Fig.~\ref{spectrumofall}.

The mass spectrum of $\ps_1$ has a sizable spike at mildly tachyonic values: $m_1^2 \gtrsim -0.1 H^2$.  
The lower bound on the mass-squared is a consequence of conditioning on prolonged inflation: at a {\it{generic}} point in the field space, not along an inflationary trajectory, $\ps_1$ would have a much stronger tachyonic instability.
In contrast, $\ps_2$ is almost never tachyonic: we found $m_2^2 < 0$ in only 4 out of 18731 realizations.  Moreover, $\ps_3$ through $\ps_6$ were never tachyonic in our ensemble.

An important property of the spectrum is that $m_3^2 \approx m_4^2$ and $m_5^2 \approx m_6^2$, to an accuracy of a few percent.  This is not a statistical statement: these eigenvalues are degenerate in each realization, not just after taking the ensemble average.
The corresponding histograms are overlaid in  Fig.~\ref{spectrumofall}.

Although we have evaluated the masses at Hubble exit, each mass  slowly changes during the course of inflation, in a  predictable way.
We find that the four heaviest fields become fractionally more massive: in one e-fold around Hubble crossing, there is a change $\delta m_a^2 \sim .02 m_a^2$ for $a=3\ldots 6$. In contrast, the two lightest fields diminish in mass, with $\delta m_2^2 \sim -.01 m_2^2$ and $\delta m_1^2 \sim -0.18 |m_1|^2$.

The masses that are relevant for determining the evolution of the fluctuations consist of a standard part coming from the potential, namely the  $V_{; IJ}$ term in equation (\ref{masssquared}),
as well as kinematic contributions involving time derivatives of background quantities, corresponding to the remaining terms in  equation (\ref{masssquared}).
We find that the kinematic contributions introduce corrections at the level of one part in $10^4$, and so can be neglected for practical purposes.

\subsection{A random matrix model for the masses} \label{rmt}

We will now show that distinctive qualitative features of the mass spectrum can be understood using random matrix theory.
The framework for this analysis is `random supergravity', by which we mean an ensemble of four-dimensional ${\cal{N}}=1$ supergravity theories whose K\"ahler potential and superpotential are random functions, in a sense made precise in \cite{MMW}.
A matrix model that is slightly simpler than that of \cite{MMW} will suffice for our purposes: we take the Hessian matrix ${\cal H}$ to be of the form\footnote{Cf.~\cite{MMW} for an explanation of the relationship between this model and the full Hessian matrix in supergravity.  We thank T.~Wrase for helpful discussions of this point.}
\begin{equation} \label{abc}
{\cal H} =
\left(
\begin{array}{c c}
A\bar A + B\bar B&  C \\
 \bar C & \bar A A +  \bar B B \end{array}
\right)\,,
\end{equation} up to a shift proportional to the identity matrix, where $A$, $B$, and $C$ are $3\times 3$ complex symmetric matrices.  We take the entries of $A$, $B$, and $C$ to be random complex numbers drawn from a normal distribution; as explained at length in \cite{MMW}, the spectrum of ${\cal H}$ is essentially independent of the statistical properties of the matrix entries.  For $C=0$ the spectrum is positive-definite and doubly degenerate, but $C \neq 0$ breaks the degeneracies and permits negative eigenvalues.

\begin{figure}[!h]
  \begin{center}
    \includegraphics[width=5.0in,angle=0]{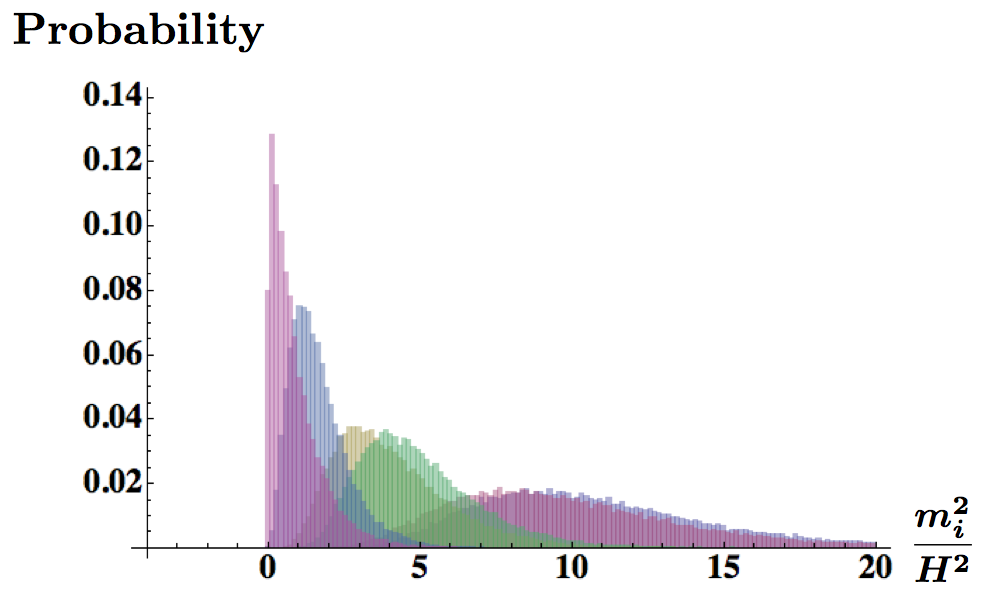}
    \caption{The mass spectrum simulated in the matrix model given in equation (\ref{abc}), cf.\ \cite{MMW}.  Comparing to Fig.~\ref{spectrumofall}, we see that the model is in good qualitative agreement with the results of simulations in the ensemble of inflaton potentials.}
    \label{spectrumRMT}
  \end{center}
\end{figure}

By requiring inflation, one has effectively imposed a lower bound on $m_1^2$, which clearly affects the empirical spectrum shown in  Fig.~\ref{spectrumofall}.  We should therefore impose a corresponding restriction in the matrix model: we take $m_1^2 \ge -0.1 H^2$.  (The overall scale is arbitrary in the matrix model, and we set our units by matching the right tail of the rightmost peak in Fig.~\ref{spectrumofall}.)  In Fig.~\ref{spectrumRMT} we show the results of simulations of the spectrum in this simple matrix model.  By adjusting parameters --- such as the weights assigned to $A$, $B$, and $C$, or the relative variance of their entries --- one can achieve reasonably good quantitative modeling of the empirical spectrum, but it is not clear that this is physically meaningful.  We have instead presented the results for the simplest case, with equal weights for all matrices, and equal variance for all entries, for which the qualitative agreement is already surprisingly good.  Notice that the eigenvalues are approximately pairwise degenerate, which matches the empirical result for
$m_3^2 \approx m_4^2$ and $m_5^2 \approx m_6^2$ but is a less accurate model of $m_1^2$ and $m_2^2$.
In contrast to the analysis of \cite{MMW}, the matrices in question are not large: they are $3\times 3$, so that large $N$ arguments (where $N$ is the size of the matrix) are marginal at best.  However, for $C=0$ and any $N$, the spectrum of ${\cal H}$ is exactly doubly degenerate, and the approximate degeneracies in the mass matrix with $C \neq 0$ hold at finite $N$.

Another key feature of the empirical spectrum that is reproduced in the matrix model is the sharp left edge in the probability density for $m_1^2$.  The requirement of $N_e \gtrsim 66$ e-folds of inflation effectively imposes  a lower bound $m_1^2 \gtrsim -0.1 H^2$.  The steep pileup of the probability density near this edge is a characteristic feature of eigenvalue spectra  of ensembles of random matrices subject to constraints on the smallest eigenvalues.

We conclude that qualitative features of the mass spectrum arising from the scalar potential on the conifold can be reproduced in the very general random matrix model of \cite{MMW}, or its simplified version (\ref{abc}), which governs any ${\cal N}=1$ supergravity theory for which the K\"ahler potential and superpotential are random functions to good approximation.
Correspondingly, these features are plausibly properties of a general random supergravity theory, not consequences of the particular conifold setting of the present work.
This strongly motivates using kindred matrix models to study much more general many-field inflationary scenarios.

To make a further observation about the scope of the matrix model, we briefly recall the method used in \cite{Baumann10} to compute the scalar potential for a D3-brane on the conifold.
The computation of \cite{Baumann10}  amounted to determining the most general solutions of ten-dimensional supergravity, in expansion around the Klebanov-Strassler solution, with certain asymptotics.  As such, these solutions incorporate in full detail the structure of a conifold geometry attached to a stabilized compactification, but --- being intrinsically ten-dimensional --- lack any manifest four-dimensional ${\cal N}=1$ supersymmetry.
In particular, the K\"ahler potential and superpotential of the four-dimensional theory arising upon dimensional reduction were not obtained directly in \cite{Baumann10}.
We therefore find it remarkable that a random matrix model based on four-dimensional random supergravity is in excellent agreement with our calculation of the mass spectrum arising from the ten-dimensional results of \cite{Baumann10}.

A further quantity of significant interest is the number of scalar fields that are light enough to fluctuate during inflation. We will now show that an efficient measure of the number of fluctuating scalars is a weighted average that takes into account the reduced contributions of fields with masses $m \to 3/2 H$ that barely fluctuate.  To determine the proper weighting, we consider the equation of motion for the perturbations.
The canonically normalized
perturbation $v=a Q$ of a test scalar field of mass $m$ in de Sitter spacetime obeys the equation of motion, in Fourier space,
\begin{eqnarray}
v_k'' + \left( k^2 - \frac{1}{\tau^2}\left(\nu^2-\frac{1}{4}\right) \right) v_k = 0 \,,
\label{eq:nu}
\end{eqnarray}
where
\begin{equation}
\nu^2 \equiv \frac{9}{4}-\frac{m^2}{H^2}\,,
\end{equation}
and $'$ denotes a derivative with respect to conformal time $\tau$. The solution of this equation with the appropriate Bunch-Davies behavior inside the Hubble radius reads
\begin{equation}
v_k(\tau)=\frac{\sqrt{\pi}}{2}e^{i(\nu+1/2)\pi/2}\sqrt{-\tau}\, H_{\nu}^{(1)}(-k\tau)\,,
\label{sol-Hnu}
\end{equation}
where $H_{\nu}^{(1)}$ is the Hankel function of the first kind, of order $\nu$, and we use the convention
\begin{eqnarray}
\nu&=& \sqrt{\nu^2} \, \,\, \, \,\,\, \,\,\,{\rm when}\, \,\nu^2 > 0\,, \\
\nu&=& i \sqrt{-\nu^2} \,\,\,{\rm when}\,\, \nu^2 < 0\,.
\end{eqnarray}
The power spectrum of fluctuations at Hubble crossing, i.e.\ for $-k \tau_\star=1$, can then be computed as
\begin{eqnarray}
{\cal P}_{Q_k,m^2}(\tau_\star) \equiv \frac{k^3}{2 \pi^2} |Q_k(\tau_\star)|^2\,.
\label{Power-Q}
\end{eqnarray}
For each of our inflationary realizations, one can sum the six contributions of the form (\ref{Power-Q}) corresponding to the six eigenvalues of the Hessian matrix determined in \S\ref{spectrumobservations}. We thus obtain an analytical estimate of the total power spectrum of scalar field fluctuations at Hubble exit,
\begin{eqnarray}
{\cal P}_{{\rm tot}} = \sum_{i=1}^6  {\cal P}_{Q_k,m_i^2}(\tau_\star)\,.
\label{Power-Qest}
\end{eqnarray}
In realizations with small bending of the trajectory (see \S\ref{turningsection}), the estimate (\ref{Power-Qest}) agrees extremely well with a direct numerical calculation of the total power spectrum of field fluctuations at Hubble exit, with a precision of order 0.01\%. In more general models,
the estimate (\ref{Power-Qest}) is only qualitatively correct, and our full numerical treatment is necessary (see for instance \S\ref{destructive}).

\begin{figure}[!h]
  \begin{center}
    \includegraphics[width=4.0in,angle=0]{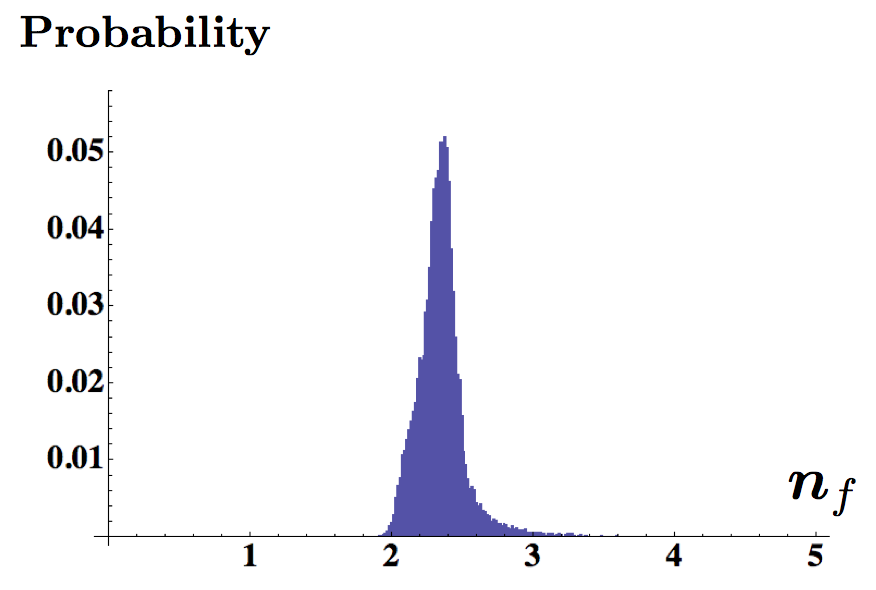}
    \caption{Histogram showing the relative probability of the number $n_f$ of scalar fields that are light enough to fluctuate during inflation, cf.\ equation (\ref{nf}).}
    \label{Hankelargument}
  \end{center}
\end{figure}

The effective number of fluctuating fields, $n_{f}$, can then be defined by comparing the estimate (\ref{Power-Qest}) to the corresponding expression for a {\it{massless}} scalar field:
\begin{eqnarray}
n_f \equiv {\cal P}_{{\rm tot}} /   {\cal P}_{Q_k,m^2=0}(\tau_\star)\,.
\label{nf}
\end{eqnarray}
In Fig.~\ref{Hankelargument} we show a histogram of  $n_f$ in our ensemble of inflationary realizations. Remarkably, the histogram is sharply peaked, and $n_f$ falls between $2$ and $3$ in about 99\% of our realizations.

We caution the reader that the number of fields that are light enough to fluctuate, $n_{f}$, is in general different from the number of fields  $n_{{\cal{R}}}$ that contribute to the curvature perturbation: for example, if the background trajectory is straight, then entropic perturbations are not converted to curvature perturbations, and $n_{{\cal{R}}}=1$ for any $n_{f}$.  The calculation above concerns the mass spectrum, and hence determines the statistical distribution of $n_{f}$. We will determine the distribution of $n_{{\cal{R}}}$ in \S\ref{manyincidence}.

\subsection{Violations of slow roll} \label{srv}

A pivotal property of the scalar mass spectrum discussed in the preceding section is that the lightest field $\ps_{1}$ has a mass $m_1^2 \sim H^2$ in a large fraction of realizations yielding $N_e \gtrsim 66$ e-folds of inflation. Thus, the slow roll approximation is only marginally applicable.

\begin{figure}[!h]
  \begin{center}
    \includegraphics[width=4.0in,angle=0]{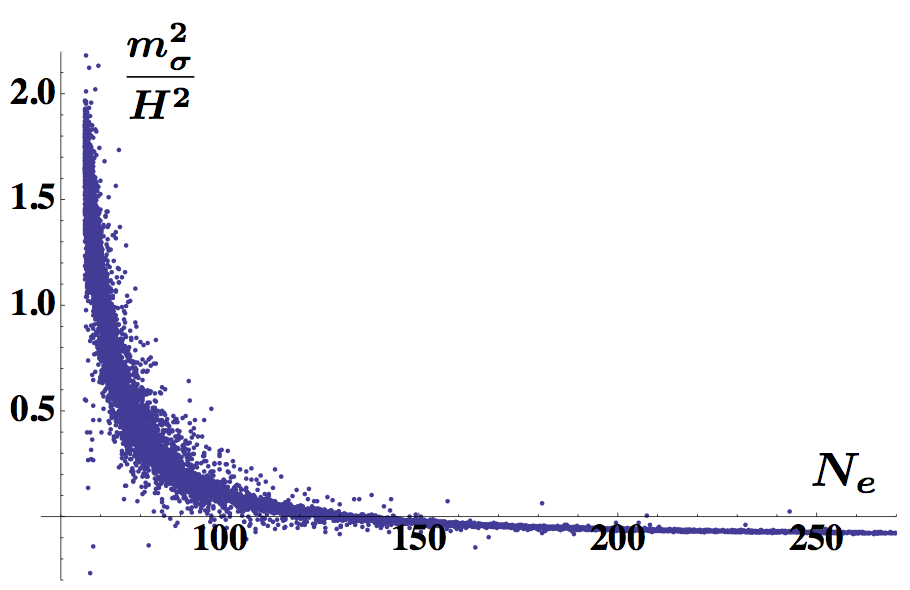}
    \caption{The mass-squared of the adiabatic fluctuation in units of $H$, evaluated 60 e-folds before the end of inflation, versus the total number of e-folds, $N_e$.}
    \label{massversusNe}
  \end{center}
\end{figure}

In Fig.~\ref{massversusNe} we show that the mass-squared of the adiabatic direction, evaluated 60 e-folds before the end of inflation, has a clear dependence on the total number of e-folds, $N_e$.  Realizations with $N_e \ll 100$ have $m_{\sigma}^2/H^2 \gtrsim 1$, while  realizations with  $N_e \gg 100$ have\footnote{It is straightforward to show analytically that 60 e-folds before the end of inflation in a single-field inflection point model yielding $N_e \gg 100$, the inflaton mass obeys $m^2 \approx -\frac{H^2}{10}$, in excellent agreement with our simulations.} $m_{\sigma}^2 \approx -0.1 H^2$. This is simply a consequence of the fact that inflation is occurring at an inflection point. Realizations yielding $N_e < 120$ have the 60 e-fold mark above the inflection point, where the curvature of the potential is positive, while realizations yielding more inflation have the 60 e-fold mark slightly below the inflection point. See \S\ref{inflection}.

\begin{figure}[h!]
\begin{center}
$\begin{array}{c@{\hspace{.1in}}c@{\hspace{.1in}}c}
\includegraphics[width=3.0in,angle=0]{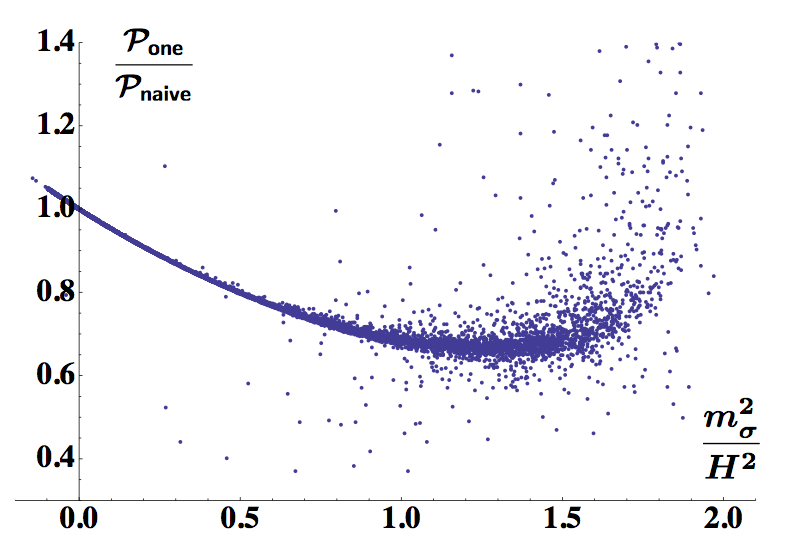} ~~~ &
\includegraphics[width=3.0in,angle=0]{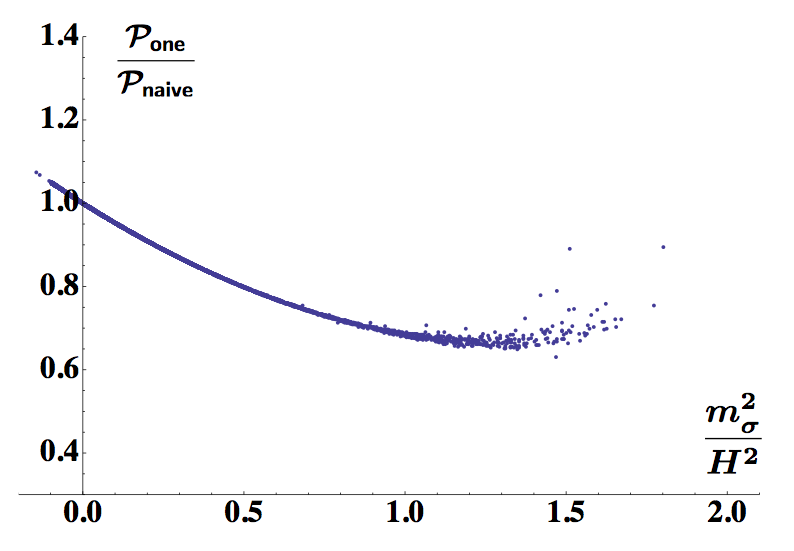}   \\
\end{array}$
\end{center}
\caption{The ratio of the {\sf one-field} power to the {\sf naive} power, versus the adiabatic mass-squared in units of $H$.
The left panel shows the entire ensemble, while the right panel is restricted to effectively single-field realizations (see \S\ref{namesofensembles}).}
\label{powerarray}
\end{figure}

Violations of slow roll  provide  significant corrections to the scalar power spectrum in our ensemble.  In Fig.~\ref{powerarray} we show the ratio of the {\sf one-field} power to the {\sf naive} power, as a function of $m_{\sigma}^2/H^2$.
We learn that increasing the mass of the adiabatic direction decreases the scalar power in a predictable manner.

\begin{figure}[!h]
  \begin{center}
    \includegraphics[width=4.0in,angle=0]{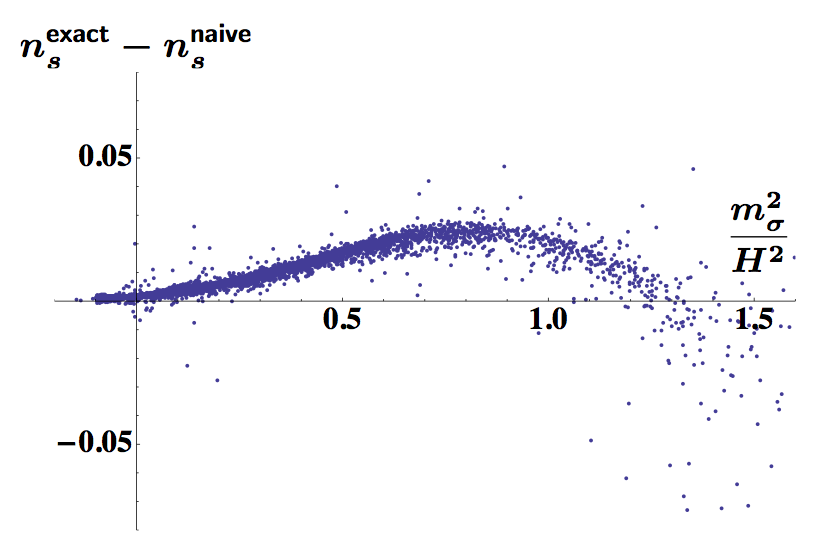}
    \caption{The difference between the {\sf exact} and {\sf naive} spectral tilts, versus the adiabatic mass-squared in units of $H$, for effectively single-field realizations (see \S\ref{namesofensembles}).}
    \label{nsadvsnsnaiveversusmass}
  \end{center}
\end{figure}

The effects of slow roll violations on the tilt for effectively single-field models are shown in Fig.~\ref{nsadvsnsnaiveversusmass}.   Except for the handful of cases with $m_{\sigma}^2/H^2 \gtrsim 1.3$, the {\sf  exact} tilt is more blue than the  {\sf naive} tilt, by an amount that is strongly correlated with $m_{\sigma}^2/H^2$.  Thus, slow roll violations tend to shift the spectrum to be slightly more blue.

\section{Multifield effects}  \label{multifield}

We now turn to our primary objective, the characterization of multifield contributions to the scalar power spectrum.
To begin, in \S\ref{isocurvature} we discuss  the degree to which multifield inflation in our ensemble is  predictive.
Next, in \S\ref{turningsection} we quantify the degree of bending of the trajectory.  In \S\ref{mincidence}  we characterize the frequency with which multifield and many-field effects arise, and in \S\ref{constraints} we describe the consequences of imposing observational constraints on the tilt.  We discuss the prospect of observable non-Gaussianities in \S\ref{NGs}.

\subsection{Decay of entropic perturbations} \label{isocurvature}
Entropic perturbations are essential for super-Hubble evolution of the curvature perturbation: if all entropic modes become massive and decay at some point after Hubble exit, the curvature perturbation becomes constant, and one says that an {\it{adiabatic limit}} has been reached (see \cite{Meyers:2010rg,Elliston:2011dr,Meyers:2011mm,Seery:2012vj} for recent discussions).
Provided that an adiabatic limit is reached during the inflating phase, one can predict the curvature perturbation on observable angular scales without knowing the details of the end of inflation or of reheating.  Conversely, failure to reach an adiabatic limit makes predictions contingent on an understanding of reheating.

The approach to an adiabatic limit clearly depends on the masses of the entropic modes: modes with $m \gtrsim H$ decay quickly outside the Hubble radius, and those with $m > \frac{3}{2}H$ do not oscillate at all.
One might a priori expect that in our ensemble all six fields have masses of order $H$, with a substantial likelihood that more than one field is tachyonic.
Our findings from  \S\ref{rmt} differ from this expectation in important details:  there is at most\footnote{Strictly speaking, we found two tachyons in a negligible fraction of trials, in 4 out of 18731 examples.} {\it{one}} tachyonic instability, the second-lightest field $\ps_{2}$ has $m_{2}^2 \sim H^2$, and the heavier fields $\ps_{3}\ldots \ps_{6}$ have masses considerably larger than $H$.  Correspondingly, we expect that the entropic modes will decay during the course of inflation,  so that an adiabatic limit is reached before the end of inflation.

\begin{figure}[!h]
  \begin{center}
    \includegraphics[width=5.0in,angle=0]{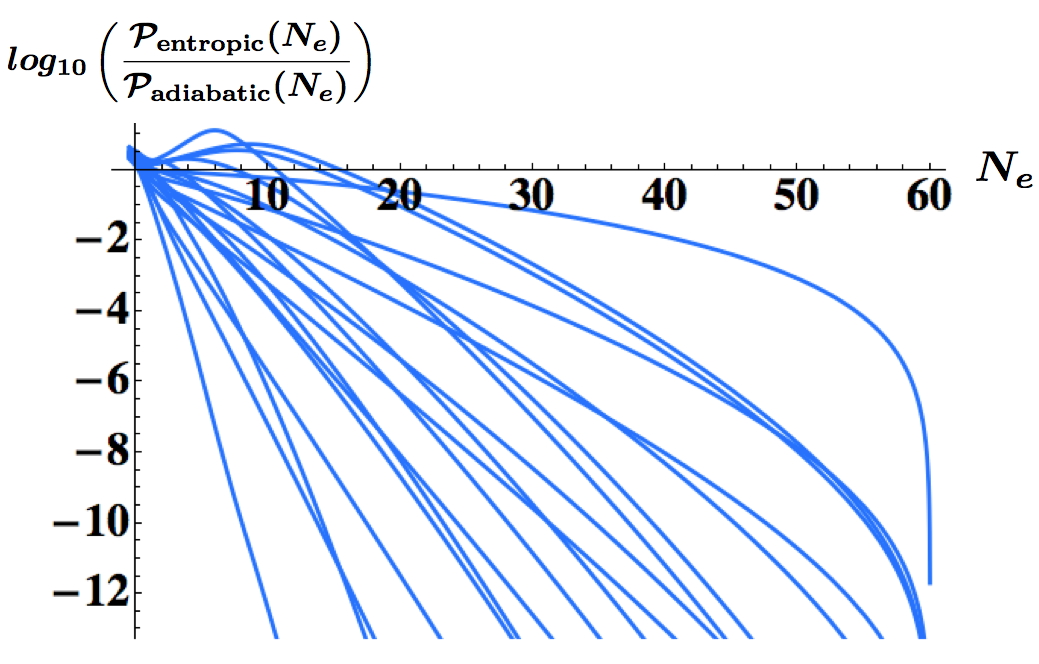}
    \caption{The curves show the rapid decay of the total power of the five entropic modes over the course of inflation, for 19 realizations.  The decaying behavior depicted is representative of the ensemble: the lightest entropic mode generically has $m^2 \gtrsim H^2$, and was tachyonic at Hubble crossing in a negligible fraction of trials.}
    \label{adiabaticlimit}
  \end{center}
\end{figure}

To see that this is the case, we examined a subset of realizations and verified that the total power of the five entropic perturbations decays exponentially.  In 60 examples, the power in entropic modes at the end of inflation was never larger than $10^{-10}$ times the adiabatic power.
In Fig.~\ref{adiabaticlimit} we show the characteristic decaying behavior for
19 examples.

%

One possible subtlety is worth mentioning.  In the argument above we have assumed that the  lightest field  $\ps_{1}$ corresponds to the adiabatic direction $\phi_{\sigma}$, so that having $m_{2}^2 \sim H^2$ and  $m_{3}^2,  m_{4}^2, m_{5}^2, m_{6}^2 \gg m_{2}^2$ ensures that  all five entropic  modes are massive enough to decay quickly.  There is good evidence for this identification: the histograms of $m_{1}^2$ and of $m_{\sigma}^2$ are nearly indistinguishable, but are quite different from those of the remaining fields.

Somewhat different results concerning the approach to the adiabatic limit in a similar system have been reported\footnote{We thank Mafalda Dias and Jonathan Frazer for discussions of this point.} by Dias, Frazer, and Liddle \cite{Dias}, who find that the angular fields have masses that are very small compared to $H$.  Correspondingly, no adiabatic limit is reached, and there is a problematic persistence of  entropic modes.  Based on the analysis of \cite{Baumann10} and \cite{Agarwal:2011wm}, or more generally on the grounds of naturalness in a theory with spontaneously broken supersymmetry, we would expect masses of order $H$ for all six fields (modulo a suppression of the mass of the adiabatic direction resulting from conditioning on prolonged inflation), which is consistent with the results reported here, but does not appear consistent with \cite{Dias}.

\subsection{Bending of the trajectory} \label{turningsection}

Because the entropic perturbations generally decay rapidly outside the Hubble radius, they are relevant for the late-time curvature perturbation only if the inflationary trajectory bends shortly after they exit the Hubble radius, so that the entropic perturbations are rapidly converted to curvature perturbations.  We therefore turn to characterizing the incidence of bending trajectories in our ensemble.

\begin{figure}[!h]
  \begin{center}
    \includegraphics[width=5.0in,angle=0]{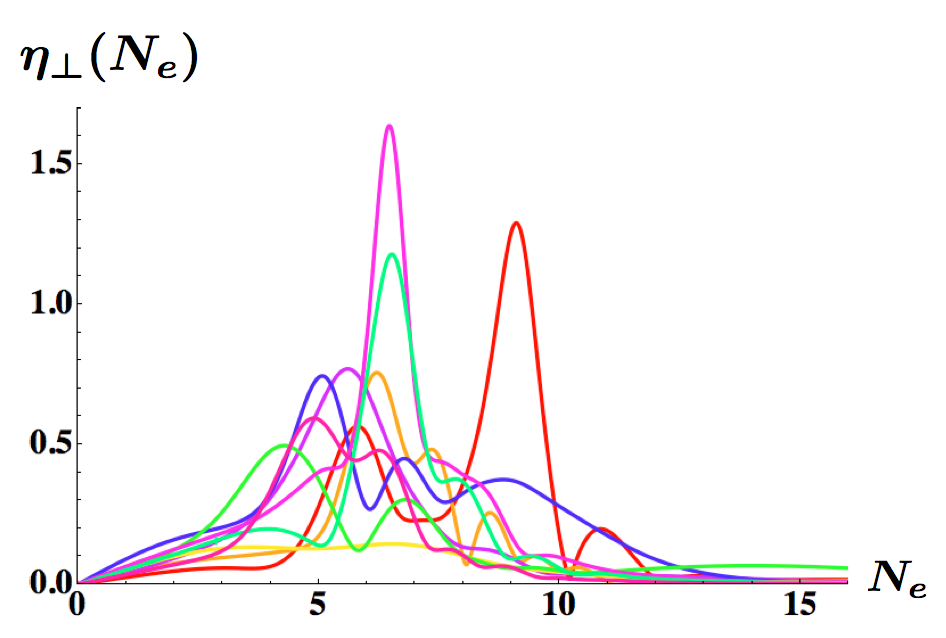}
    \caption{The curves show the rapid changes in $\eta_\perp$ during the first $15$ e-folds for 9 realizations of inflation.  Note that $N_e=0$ refers to the start of inflation in each realization, rather than to the moment when the CMB crosses the Hubble radius, which occurs much later in many examples.}
    \label{etaperp-initial}
  \end{center}
\end{figure}

A very useful measure of turning is the parameter $\eta_{\perp}$ defined in equation (\ref{etaperpdefinition}), which measures the acceleration of the trajectory transverse to the instantaneous velocity.
Analytic methods that make a slow roll approximation around the time of Hubble crossing generally require $\eta_{\perp} \ll 1$, which can be related to the `slow-turn approximation' discussed in \cite{Peterson:2010np}.
Correspondingly, effects requiring $\eta_{\perp}\gtrsim 1$ at Hubble crossing are only partially understood.  Our treatment makes no approximation, and indeed $\eta_{\perp}$ can be quite large: see
Fig.~\ref{etaperp-initial}, which shows the evolution of $\eta_{\perp}$ in the first 15 e-folds of inflation in a handful of representative
examples.
The gradual decay of $\eta_{\perp}$ can be understood from the properties of the evolution near the inflection point: while the inflaton is spiraling down to the inflection point, turning is generic, but in the immediate vicinity of the inflection point, where prolonged inflation occurs, the trajectory is quite straight.

\begin{figure}[h!]
\begin{center}
$\begin{array}{c@{\hspace{.1in}}c@{\hspace{.1in}}c}
\includegraphics[width=3.2in,angle=0]{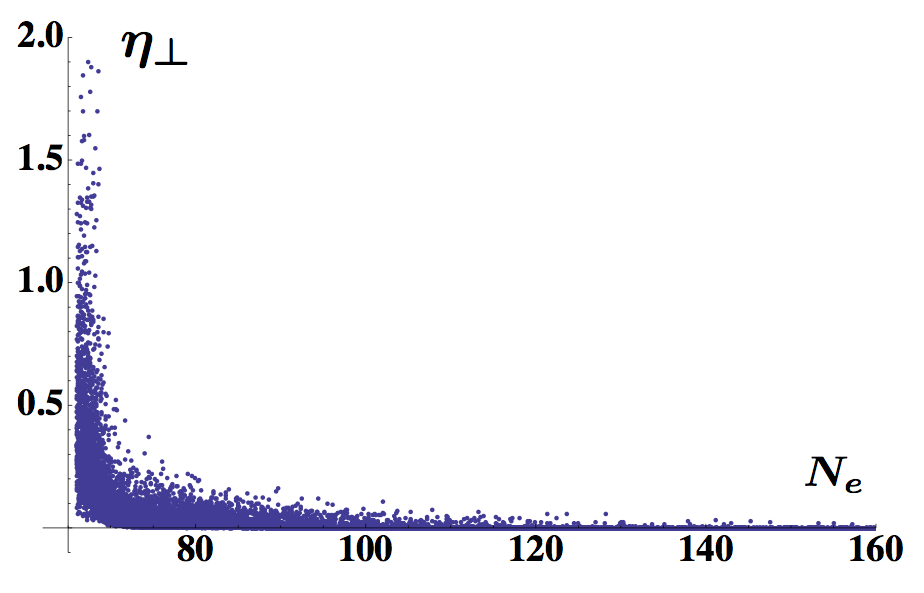} ~~~ &
\includegraphics[width=3.2in,angle=0]{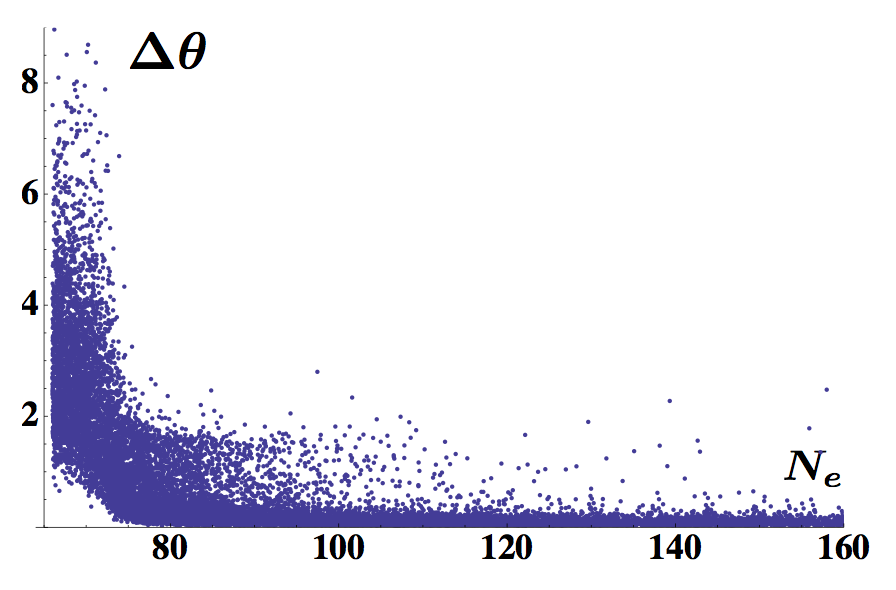}   \\
\end{array}$
\end{center}
\caption{Left panel: the slow roll parameter $\eta_{\perp}$ evaluated at Hubble crossing, versus the total number of e-folds, $N_e$.  Right panel: the total turn of the trajectory, from six e-folds before Hubble crossing until the end of inflation, $\Delta \theta \equiv\int_{HC-6}^{end} dN_e\,\eta_{\perp}$, versus the total number of e-folds, $N_e$.}
\label{turnarray}
\end{figure}

In the left panel of Fig.~\ref{turnarray} we plot $\eta_{\perp}$, evaluated at Hubble crossing, versus the total number of e-folds, $N_e$. Evidently, $\eta_{\perp}$ at Hubble crossing is quite small in scenarios giving rise to $N_e \gg 60$ e-folds of inflation, but is significant in scenarios yielding less inflation.  This is consistent with the picture described above in which there are large transient contributions to  $\eta_{\perp}$ that decay after prolonged inflation.

Another practical measure of the amount of turning is the `total turn' $\Delta \theta$, which we define to be the integral of $\eta_{\perp}$ from six e-folds before Hubble crossing until the end of inflation: $\Delta \theta \equiv\int_{HC-6}^{end} dN_e\,\eta_{\perp}$.\footnote{Turns occurring shortly before Hubble crossing can have an effect on the perturbations, and we therefore define the total turn as the integral of $\eta_{\perp}$ from six e-folds before Hubble crossing to the end of inflation.}
From the right panel of Fig.~\ref{turnarray} we see that the total turn
is quite large in cases with $N_e \lesssim 80$ e-folds of inflation.
Notice that even in cases with arbitrarily many e-folds, a total turn of order $\Delta \theta \sim 1/2$ remains likely.  This is an effect from the end of inflation: as the inflaton falls off the inflection point, its trajectory very often bends (see \cite{Frazer:2011tg} for a discussion of related points).  We have verified this by checking that the restricted integral $\int_{HC-6}^{end-10} dN_e\eta_{\perp}$ is small in cases with  $N_e \gg 60$ e-folds,
even though $\int_{HC-6}^{end} dN_e\,\eta_{\perp}$ is not.

\subsection{Multifield effects and many-field effects} \label{mincidence}

Having understood the properties of the mass matrix and of the background evolution that influence the evolution of entropic perturbations, we are in a position to understand multifield contributions to the primordial perturbations.

One of the most interesting questions for a model of inflation with more than two fields is whether the primordial perturbations are effectively governed by only two fields, or instead have distinctive `many-field' signatures (see \cite{Peterson:2011yt} for a recent discussion).  In a general $N$-field model of inflation, as we reviewed in \S\ref{decomposition}, one of the $N-1$ entropic fluctuations, which we call the first entropic fluctuation, plays a distinguished role by instantaneously coupling to the adiabatic perturbation. Most explicit studies of inflation with more than one field have taken $N=2$ for simplicity, in which case the entropic subspace is one-dimensional and this distinction is unnecessary.
However, for our system there are five entropic modes, each of which can in principle contribute to the curvature perturbation.\footnote{Although the second through fifth entropic modes do not couple instantaneously to the adiabatic perturbation, they do couple to the first entropic mode, which can then source the adiabatic perturbation when the trajectory bends.}
We must therefore study not just the incidence of multifield effects, but also the incidence of `many-field'\footnote{For the purpose of this discussion, because `multifield inflation' means `inflation with two or more light fields', we will use `many-field inflation' to refer to `inflation with three or more light fields'.  An important and challenging problem is to understand inflation with $N \gg 1$ light fields, but we do not use `many' in this sense in the present work.} effects, in which the curvature perturbation is not dictated by the adiabatic and first entropic fluctuations alone, and instead receives contributions from the higher (second through fifth) entropic modes. The method for this analysis was described in \S\ref{pertmethod}: when ${\cal P}_{{\sf exact}} \neq {\cal P}_{{\sf k}}$, then at least $k+1$ modes contribute to the curvature perturbation.

\subsubsection{Illustrative examples}

\begin{figure}[h!]
\begin{center}
\includegraphics[width=6.5in,angle=0]{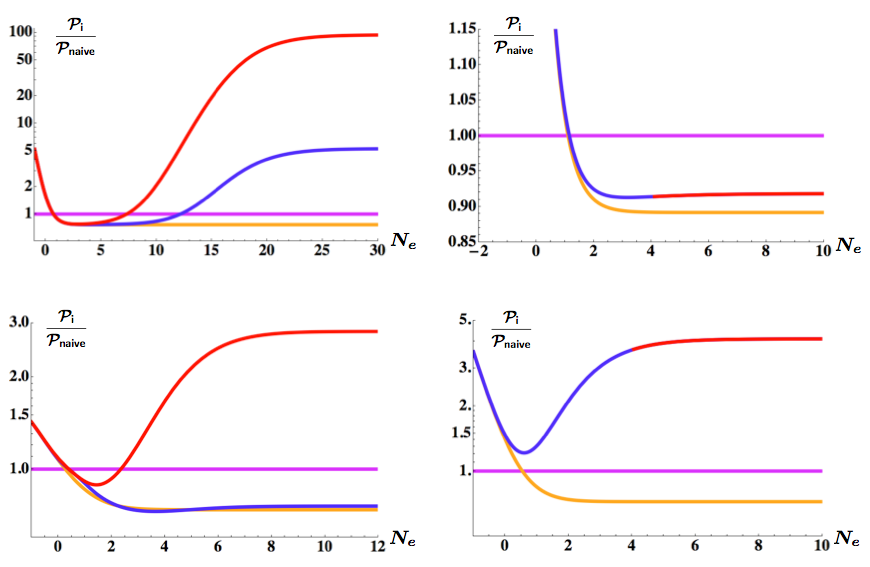}
\caption{Scalar power spectra as functions of the number of e-folds since Hubble crossing, for four different realizations. The purple, orange, blue, and red lines correspond to the {\sf naive}, {\sf one-field}, {\sf two-field}, and {\sf exact} spectra, respectively.  Lines with combined colors indicate that the corresponding spectra overlap to high accuracy.  Upper left: many-field and two-field effects are both important.  Upper right: two-field effects are present, but many-field effects are negligible.  Lower right: two-field effects are present but many-field effects are negligible, in a rare example that is consistent with constraints on the tilt (see \S\ref{constraints}).  Lower left: two-field effects are negligible, but many-field effects are large.} \label{evolvearray}
\end{center}
\end{figure}

We begin by examining a collection of illustrative examples, shown in Fig.~\ref{evolvearray}. For each example we display the power spectrum as a function of the number of e-folds since Hubble crossing, using the {\sf naive}, {\sf one-field}, {\sf two-field}, and {\sf exact} models.

In the upper left panel we show a striking example for which all four models give different results.  Around Hubble crossing the various models give roughly comparable results for  the curvature perturbation, but conversion of entropic perturbations to curvature perturbations after Hubble crossing causes the {\sf two-field} and {\sf exact} results to grow sharply.  As ${\cal P}_{{\sf exact}} \neq {\cal P}_{{\sf one}}$, multifield effects cannot be ignored, but because also ${\cal P}_{{\sf exact}} \neq {\cal P}_{{\sf two}}$, we see that a two-field description is inadequate.  Notice the scale: many-field contributions to the curvature perturbation dwarf the two-field contribution, which is itself non-negligible.

Next, in the upper right panel of Fig.~\ref{evolvearray} we exhibit a case in which  ${\cal P}_{{\sf exact}} \approx {\cal P}_{{\sf two}} \neq {\cal P}_{{\sf one}}$, so that two-field effects cannot be omitted, but many-field effects are negligible.  The lower left panel shows an example in which  ${\cal P}_{{\sf exact}} \neq {\cal P}_{{\sf two}} \approx {\cal P}_{{\sf one}}$, for which many-field effects are significant but two-field effects are negligible.
Finally, the lower right panel shows a rare example in which ${\cal P}_{{\sf exact}} \approx {\cal P}_{{\sf two}} \neq {\cal P}_{{\sf one}}$, so that two-field but not many-field effects are important, and the resulting power spectrum is compatible with observational constraints on the tilt.

\subsubsection{The incidence of multifield effects}  \label{incidence}

\begin{figure}[!h]
  \begin{center}
    \includegraphics[width=5.0in,angle=0]{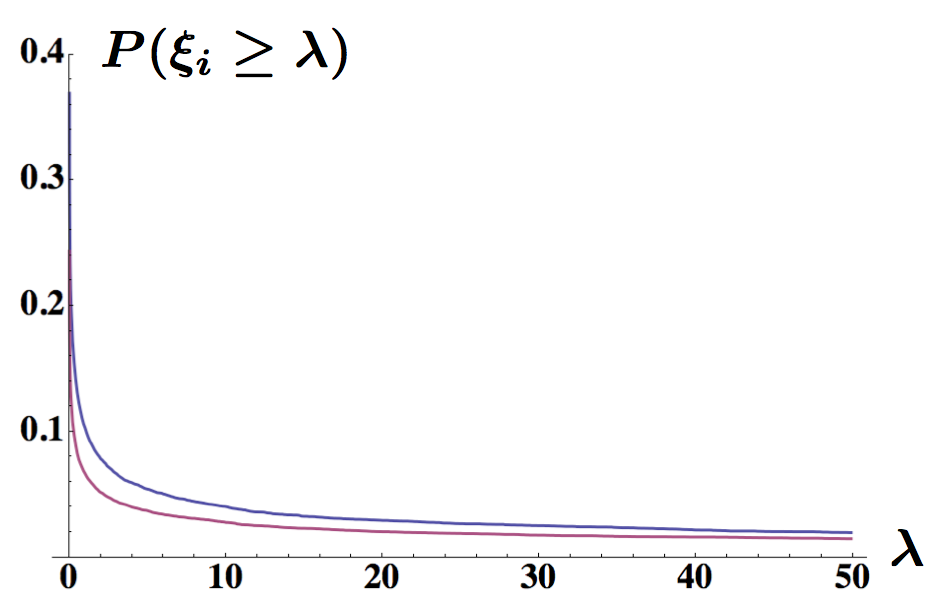}
    \caption{The relative probability $P(\xi \ge \lambda)$  of different levels of corrections of the scalar power.
    The upper curve shows corrections due to multifield effects,  corresponding to $\xi_{{\sf exact}/{\sf one}} \equiv |{\cal P}_{{\sf exact}}/{\cal P}_{{\sf one}}-1|$.
    The lower curve shows corrections due to many-field effects, corresponding to $\xi_{{\sf exact}/{\sf two}} \equiv |{\cal P}_{{\sf exact}}/{\cal P}_{{\sf two}}-1|$.
    Note that $\lambda=1$ corresponds to a 100\% correction, and the tails extend to very large enhancements.}\label{quantilemultifield}
  \end{center}
\end{figure}

Equipped with a few examples of the possible phenomenology, we now proceed to a statistical analysis of the incidence of multifield effects in our ensemble.
First, in Fig.~\ref{quantilemultifield} we show the cumulative probability that multifield contributions to the scalar power have a given size.  The upper curve corresponds to $\xi_{{\sf exact}/{\sf one}} \equiv |{\cal P}_{{\sf exact}}/{\cal P}_{{\sf one}}-1|$.
We find that approximately  30\% of realizations have  $\xi_{{\sf exact}/{\sf one}} \ge .01$, corresponding to a 1\% correction to the power, and 10\% of realizations have $\xi_{{\sf exact}/{\sf one}} \ge 1$, corresponding to a 100\% correction to the power.
The tail of the distribution is significant even at very large enhancements.

\begin{figure}[!h]
  \begin{center}
    \includegraphics[width=5.0in,angle=0]{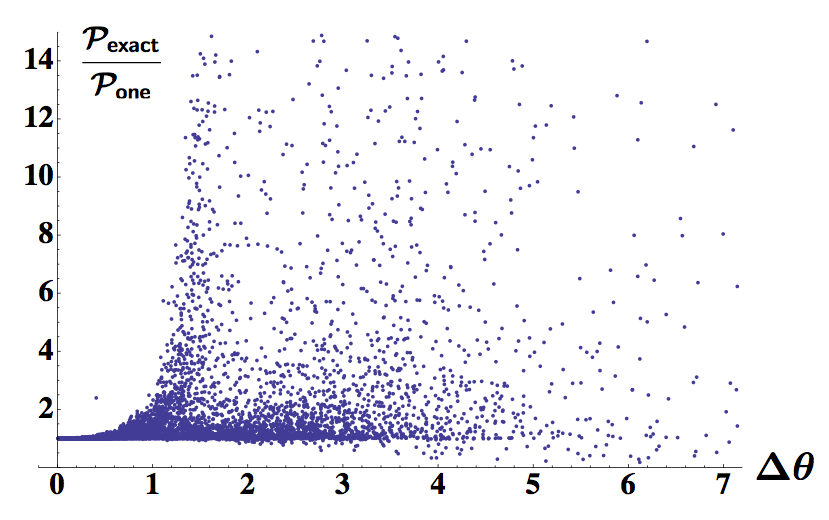}
    \caption{The ratio of the {\sf exact} power to the {\sf one-field} power, versus the total turn of the trajectory during inflation, for all realizations.  Notice the clear threshold of total turning required to achieve a given  ratio of multifield to single-field power.} \label{totalturn1}
  \end{center}
\end{figure}

Next, in  Fig.~\ref{totalturn1} we show the relationship between the multifield correction to the power spectrum and the total turn $\Delta \theta$ of the inflationary trajectory.
An important fact visible in Fig.~\ref{totalturn1} is that for any value of the enhancement in power due to multifield effects, ${\cal P}_{{\sf exact}}/{\cal P}_{{\sf one}}$, there is a minimum amount of total turn $\Delta \theta$ necessary to effect such an enhancement.  For total turning $\Delta \theta$ exceeding a threshold value $\Delta \theta_{\star} \approx 1.4$, the enhancement in power can be very large.  For practical purposes, the vertical range of Fig.~\ref{totalturn1} has been restricted; see Fig.~\ref{quantilemultifield} for an alternative representation of the incidence of large enhancements.

\subsubsection{The incidence of many-field effects}  \label{manyincidence}

We now turn to characterizing the incidence of many-field effects, characterized by $\xi_{{\sf exact}/{\sf two}} \equiv |{\cal P}_{{\sf exact}}/{\cal P}_{{\sf two}}-1|$.  We remind the reader that the {\sf two-field} truncation retains the instantaneous adiabatic and first entropic modes, so that $\xi_{{\sf exact}/{\sf two}} > 0$ indicates that additional entropic modes contribute to the perturbations.
For practical reasons, although we computed the {\sf exact} scalar power for all 18731 realizations yielding at least 66 e-folds of inflation, we computed the {\sf two-field} power
only in the effectively multifield ensemble, for which $\xi_{{\sf exact}/{\sf one}} \equiv |{\cal P}_{{\sf exact}}/{\cal P}_{{\sf one}}-1| \ge 0.01$.  The logic is that many-field effects will be significant only if multifield effects are significant, so that we can compute the frequency of many-field effects in the full ensemble without computing ${\cal P}_{{\sf two}}$ in the effectively single-field subset.\footnote{In principle, one can envision a scenario in which ${\cal P}_{{\sf exact}} \approx {\cal P}_{{\sf one}} \ll {\cal P}_{{\sf two}}$, so that {\sf  two-field} and many-field effects are both present but cancel out in the {\sf exact} power (see \S\ref{destructive} for a discussion of this point).  Correspondingly, some scenarios with $\xi_{{\sf exact}/{\sf one}} < 0.01$ could involve significant  many-field effects. We have verified that this occurrence is very infrequent in our ensemble, occurring in less than 1\% of realizations.}

\begin{figure}[!h]
  \begin{center}
    \includegraphics[width=5.0in,angle=0]{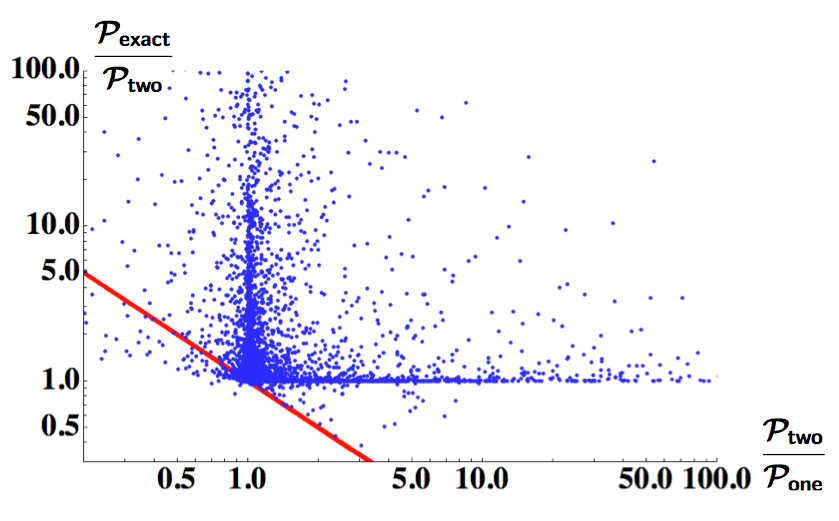}
    \caption{Many-field and two-field effects, for effectively multifield models  (see \S\ref{namesofensembles}).
    Two distinct populations are visible: the vertical branch has negligible two-field effects and significant many-field effects, while the horizontal branch has significant two-field effects and negligible many-field effects.  Points below the red line have destructive multifield effects, i.e. ${\cal P}_{{\sf exact}} < {\cal P}_{{\sf one}}$, cf.~\S\ref{destructive}.} \label{2fieldmanyfield}
  \end{center}
\end{figure}

Many-field effects are only slightly less common than multifield effects in general: many-field corrections exceed 1\% (i.e., $\xi_{{\sf exact}/{\sf two}} \ge 0.01$) in 18\% of all realizations, and exceed 100\%
in 6\% of realizations.  Among effectively multifield models, many-field effects are commonplace: $\xi_{{\sf exact}/{\sf two}} \ge 0.01$ in 62\% of models with $\xi_{{\sf exact}/{\sf one}} \ge 0.01$.
The lower curve in Fig.~\ref{quantilemultifield} indicates the cumulative probability that many-field contributions to the scalar power have a given size, as measured by $\xi_{{\sf exact}/{\sf two}}$.  Comparing to the upper curve, which corresponds to $\xi_{{\sf exact}/{\sf one}} \equiv |{\cal P}_{{\sf exact}}/{\cal P}_{{\sf one}}-1|$, we see that when multifield effects are large, the likelihood of many-field effects grows dramatically.
On the other hand, large many-field effects and large two-field effects are rarely present in the same model.  The anticorrelation between two-field and many-field effects is clearly visible in Fig.~\ref{2fieldmanyfield}, which primarily consists of two distinct populations, one with exclusively two-field effects (horizontal branch) and another with exclusively many-field effects (vertical branch).  It is worth stressing the peculiarity of this latter population: based on a {\sf two-field} description, one would wrongly conclude that these models have negligible multifield effects, while in fact only higher entropic modes affect the curvature perturbation.

\subsubsection{Destructive multifield effects} \label{destructive}

An interesting phenomenon visible in Fig.~\ref{totalturn1} and Fig.~\ref{2fieldmanyfield} is destructive interference, in which ${\cal P}_{{\sf k+1}} < {\cal P}_{{\sf k}}$ for some $k$.
Destructive two-field effects are not uncommon, but one can see from Fig.~\ref{2fieldmanyfield} that the majority of cases with ${\cal P}_{{\sf two}} < {\cal P}_{{\sf one}}$ have ${\cal P}_{{\sf exact}} > {\cal P}_{{\sf one}}$, so that the net effect of all entropic modes is to add power.  However, a handful of examples with ${\cal P}_{{\sf exact}} < {\cal P}_{{\sf one}}$ are visible in Fig.~\ref{totalturn1}.

In certain approximate treatments of the evolution of the perturbations in multifield inflation, the addition of entropic modes can only increase the power spectrum of the curvature perturbation $\cal R$, and the destructive multifield effects that we observe cannot arise.  We will therefore pause to explain why destructive multifield effects are not forbidden, and what implicit assumptions lead to the conclusion that destructive multifield effects cannot arise.
This discussion is somewhat decoupled from the rest of the paper and can be omitted in a first reading.

Let us consider a two-field inflationary model for the sake of simplicity.  In the {\sf exact}
description, a possible quantization scheme consists of writing
\begin{eqnarray}
\hat \R^{\sf two}_{\ka} = \R^{\sigma}_k \hat a^{\sigma}_{\ka} +  (\R^{\sigma}_k)^* (\hat a^{\sigma}_{-\ka} )^\dagger  + \R^{s}_k \hat a^{s}_{\ka} +  (\R^{s}_k)^* (\hat a^{s}_{-\ka} )^\dagger\,,
\label{quantization-2}
\end{eqnarray}
where the creation and and annihilation operators satisfy
\begin{eqnarray}
[\hat a_{\ka}^I,(\hat a_{\ka'}^J)^{\dagger}]=\delta^{IJ} \delta(\ka-\ka')\,,
\label{commutation}
\end{eqnarray}
with $I,J \in (\sigma,s)$. The connection to the method of numerical evolution described in \S\ref{pertmethod} is as follows. The exact coupled equations of motion are solved twice, for two different initial conditions. In a first run, the adiabatic fluctuation begins in the Bunch-Davies vacuum, while the entropic perturbation is initially set to zero. The corresponding solution for $\cal R$ leads to $ \R^{\sigma}_k$. In a second run, the entropic fluctuation begins in the Bunch-Davies vacuum, while the adiabatic perturbation is initially set to zero. The corresponding solution for $\cal R$ leads to $ \R^{s}_k$. The total power spectrum of ${\cal R}$ is then given by
\begin{eqnarray}
 \langle 0| \hat \R^{\sf two}_{\ka}  \hat \R^{\sf two}_{\ka'}   |0\rangle = \delta({\bf k} + {\bf k}')  \left( |\R^{\sigma}_k|^2 + |\R^{s}_k|^2  \right)\,.
 \label{result-2}
\end{eqnarray}
Thus, in the {\sf  exact} description, there are two contributions to the curvature perturbation that add in quadrature.

In the {\sf one-field} description, on the other hand, one simply writes
\begin{eqnarray}
\hat \R^{\sf one}_{\ka}&=& \R_k \hat a^{\sigma}_{\ka} +  (\R_k)^* (\hat a^{\sigma}_{-\ka} )^\dagger \,,
\label{quantization-1}
\end{eqnarray}
and the entropic fluctuations are set to zero for all time, which leads to
\begin{eqnarray}
 \langle 0| \hat \R^{\sf one}_{\ka}  \hat \R^{\sf one}_{\ka'}   |0\rangle = \delta({\bf k} + {\bf k}')  |\R_k|^2 \,.
\end{eqnarray}
If $\R^{\sigma}_k$ in equation \refeq{quantization-2} and $\R_k$ in equation \refeq{quantization-1} were identical, then we would conclude that entropic perturbations affect the two-point function of the curvature perturbation exclusively through the term $|\R^{s}_k|^2$ in equation \refeq{result-2}, and this contribution is manifestly nonnegative.

In fact, however, $\R^{\sigma}_k$ in equation \refeq{quantization-2} and $\R_k$ in equation \refeq{quantization-1} are quite different.
In the evolution that determines $\R^{\sigma}_k$, the entropic fluctuations are initially set to zero, but are not forced to be zero for all time: entropic fluctuations can be generated through coupling with the adiabatic mode, and can then backreact on the adiabatic mode itself.
Thus, it can happen that $|\R^{\sigma}_k|^2 < |\R_k|^2$, and in turn it is logically possible --- though not necessarily common --- to have
$|\R^{\sigma}_k|^2 + |\R^{s}_k|^2  < |\R_k|^2 $, i.e.\ ${\cal P}_{{\sf two}} < {\cal P}_{{\sf one}}$.

Another perspective on this effect comes from considering the coupling between the adiabatic and entropic perturbations during Hubble crossing.
A convenient assumption that is often utilized (but not always explicitly invoked) in the literature is that $\eta_\perp \ll 1$ during the few e-folds of Hubble crossing, so that the adiabatic and entropic perturbations evolve independently during that time.
In this approximation, adiabatic and entropic perturbations are uncorrelated soon after Hubble crossing, say at a time $t_\star$ at which spatial gradients can already be neglected, so that one can write (again taking a two-field model for simplicity)
\begin{eqnarray}
\hat \R^{\sf two}_{\ka}= \R^{\sigma}_k \hat a^{\sigma}_{\ka \star} +  (\R^{\sigma}_k)^* (\hat a^{\sigma}_{-\ka \star} )^\dagger  + \R^{s}_k \hat a^{s}_{\ka \star} +  (\R^{s}_k)^* (\hat a^{s}_{-\ka \star} )^\dagger\,,
\label{quantization-3}
\end{eqnarray}
where $\hat a^{\sigma}_{\ka \star}$ and $ \hat a^{s}_{\ka \star}$ satisfy the relation \refeq{commutation}, and in particular are {\it independent}.
We stress that $\hat a^{\sigma}_{\ka \star}$ and $ \hat a^{s}_{\ka \star}$ are different from $\hat a^{\sigma}_{\ka}$ and $ \hat a^{s}_{\ka}$ in equation \refeq{quantization-2}: the former are defined such that the power spectrum of $\hat{\cal R}^{{\sf two}}_k$ at $t_\star$ coincides with $  | \R^{\sigma}_k(t_\star)|^2 $.
Just as before, $ \R^{\sigma}_k(t \geq t_\star)$ is computed by solving the full system of equations, setting the entropic perturbations to zero at $t=t_\star$, while $ \R^{s}_k(t \geq t_\star)$ is computed by solving the full system of equations, setting the adiabatic perturbations to zero at $t=t_\star$.
The power spectrum is then obtained as in equation \refeq{result-2}.
However, the assumption of decoupling during Hubble crossing results in one crucial difference, as we now explain.
As is well known, although the entropic fluctuation can source the adiabatic fluctuation on super-Hubble scales, the adiabatic fluctuation does not source the entropic fluctuation in this regime \cite{Wands:2000dp}.  Hence, if the entropic fluctuation and its time derivative are set to zero at time $t_\star$, they will remain zero, and no backreaction on $\R^{\sigma}_k(t \geq t_\star)$ is possible.  Correspondingly, if multifield effects are neglected until a time $t_\star$ after which the adiabatic mode cannot source the entropic mode --- which is precisely what occurs when the slow turn approximation is made during Hubble crossing --- then $\R^{\sigma}_k$ coincides with the {\sf one-field} quantity $\R_k$ in \refeq{quantization-1}. Because of the second term in \eqref{result-2}, one then always obtains that, under these approximations, ${\cal P}_{{\sf two}} \ge {\cal P}_{{\sf one}}$. This corresponds, for example, to the prediction of the zeroth-order
transfer functions formalism.

As should be clear from the discussion above, a necessary condition for destructive multifield effects is bending of the trajectory around the time of Hubble crossing. As a consistency check, we have verified that in our realizations with destructive multifield effects, $\eta_\perp$ always has a large peak at Hubble crossing.
However, bending at Hubble exit is not a sufficient condition: a large number of our realizations have $\eta_\perp \gtrsim 1$ during Hubble crossing but do not display destructive multifield effects.

We should point out that although ${\cal P}_{{\sf exact}} \ll {\cal P}_{{\sf one}}$ occurs in our ensemble, ${\cal P}_{{\sf exact}} \ll {\cal P}_{{\sf naive}}$ does not: when ${\cal P}_{{\sf exact}} \ll {\cal P}_{{\sf one}}$, we typically find ${\cal P}_{{\sf naive}} \ll {\cal P}_{{\sf one}}$ and
${\cal P}_{{\sf exact}} ={\cal O}({\cal P}_{{\sf naive}})$.  Even so, examples with ${\cal P}_{{\sf exact}} < {\cal P}_{{\sf naive}}$ are common:
61\% of all models have $ 0.63<{\cal P}_{{\sf exact}}/{\cal P}_{{\sf naive}} < 1$.
However, none of the realizations with ${\cal P}_{{\sf exact}} < {\cal P}_{{\sf one}}$ or ${\cal P}_{{\sf exact}} < {\cal P}_{{\sf naive}}$ has a scalar tilt in the observationally allowed region (see \S\ref{constraints}).

Finally, we remark that the amplitude of primordial gravitational waves, as measured by $r$, can be slightly larger than what the naive estimate $r=16\epsilon$ would suggest.  In several examples we indeed find ${\cal P}_{\cal T}/{\cal P}_{{\sf exact}} > 16\epsilon$, with  ${\cal P}_{\cal T}$ denoting the tensor power spectrum.  It would be interesting to understand if it is therefore possible to violate the  consistency relation $r \leq -8 n_{{\cal T}}$ \cite{transfer,Bartolo:2001rt}, but
to determine this requires computing $n_{\cal T}$ very precisely, which is challenging in models with $r\ll 1$.

\subsection{Constraints from scale invariance} \label{constraints}

The results described in the preceding sections refer to the ensemble of realizations of inflation that give rise to 66 or more e-folds of expansion followed by a hybrid exit.  Such a cosmic history serves to solve the horizon, flatness, and monopole problems, but is not necessarily consistent with observations of the CMB temperature anisotropies.  In this section we study multifield corrections to the tilt (\S\ref{multifieldtilt}), and then describe the restricted ensemble of realizations consistent with WMAP7 constraints on the tilt (\S\ref{restrictedensemble}).

 \subsubsection{Multifield contributions to the tilt}  \label{multifieldtilt}

\begin{figure}[!h]
  \begin{center}
    \includegraphics[width=4.0in,angle=0]{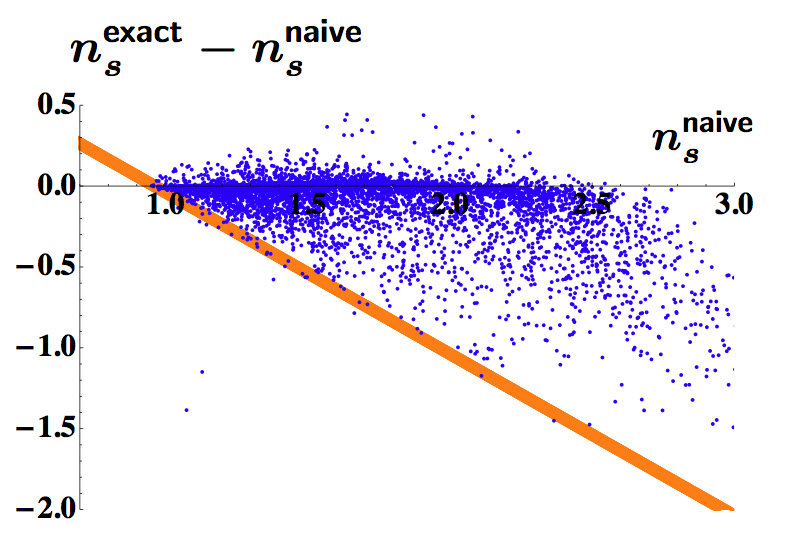}
    \caption{The difference $\delta n_s$ between the {\sf  exact} and {\sf  naive} results for the tilt $n_s$, versus the  {\sf  naive} tilt,
     for effectively multifield realizations (see \S\ref{namesofensembles}).
    Notice that multifield contributions typically shift the spectrum toward scale invariance, and the more blue the naive spectrum, the larger the scatter of $\delta n_s$.
     The orange band corresponds to the window allowed at $2\sigma$ by WMAP7.}
    \label{dnsmultifieldfigure1}
  \end{center}
\end{figure}

We begin by characterizing multifield contributions to the tilt $n_s$ of the scalar power spectrum.
In Fig.~\ref{dnsmultifieldfigure1} we show the correlation between the multifield correction to the tilt, defined as $\delta n_s \equiv n_s^{{\sf  exact}} - n_s^{{\sf  naive}}$, and the {\sf naive} tilt $n_s^{{\sf  naive}}$.
In realizations with a distinctly blue {\sf naive} spectrum, inflation is occurring well above the inflection point, and transients from the approach to the inflection point have not necessarily decayed at the time that the CMB exits the Hubble radius (see \S\ref{inflection}).  Correspondingly, there is a larger likelihood of multifield effects, as evident in Fig.~\ref{dnsmultifieldfigure1}.  We observe that multifield effects typically shift the spectrum toward scale invariance, but the size of the effect is rarely large enough to produce a model in the observational window denoted by the orange band.  There are a few very interesting cases with large multifield effects that have tilts\footnote{We have not determined whether the running of the scalar power spectrum further constrains these models.} consistent with WMAP7 at the $2\sigma$ level, which we will discuss in more detail in \S\ref{restrictedensemble}.

The reader might object that $\delta n_s \equiv n_s^{{\sf  exact}} - n_s^{{\sf  naive}}$ does not exclusively represent {\it{multifield}} effects: if  $n_s^{{\sf one}} \neq n_s^{{\sf  naive}}$ then $\delta n_s$ contains a contribution from strictly single-field slow-roll-violating effects.
Slow-roll-violating effects, for effectively single-field trajectories, were already described in \S\ref{srv} (see Fig.~\ref{nsadvsnsnaiveversusmass}).
We have verified that for the full ensemble, including realizations with large multifield effects, slow roll violations generally make small ($\delta n_s^{{\rm SR}}\lesssim 0.05$) positive contributions to $n_s$, as in the single-field case of \S\ref{srv}.
In contrast, the multifield effects described above make a much larger negative contribution to $n_s$.

\subsubsection{Realizations consistent with constraints on the tilt} \label{restrictedensemble}

In this section we present results for the restricted ensemble of realizations that are consistent with WMAP7 constraints on the {\it{tilt}} of the scalar power spectrum, which constitute 21 \% of our full ensemble. The overall amplitude of the scalar power spectrum can be adjusted by changing the scale\footnote{In the underlying microphysical model, this adjustment corresponds to changing an inflaton-independent contribution to the vacuum energy.  The magnitude of the vacuum energy could in principle be correlated with other terms in the potential, but modeling such a correlation is beyond the scope of the present work.} of the potential, and we will assume that this has been done.  We do not incorporate constraints on the running, but it could be interesting to do so.

The consequence of imposing constraints on the tilt is very striking, and easily understood by recalling the  single-field inflection point model described in \S\ref{inflection}.
A single-field inflection point model is compatible with constraints on the tilt only if the total number of e-folds is $N_e \gtrsim 120$, so that the observed CMB exits the Hubble radius when the inflaton is at or below the inflection point, at which point the spectrum becomes red.

\begin{figure}[h!]
\begin{center}
$\begin{array}{c@{\hspace{.1in}}c@{\hspace{.1in}}c}
\includegraphics[width=3.6in,angle=0]{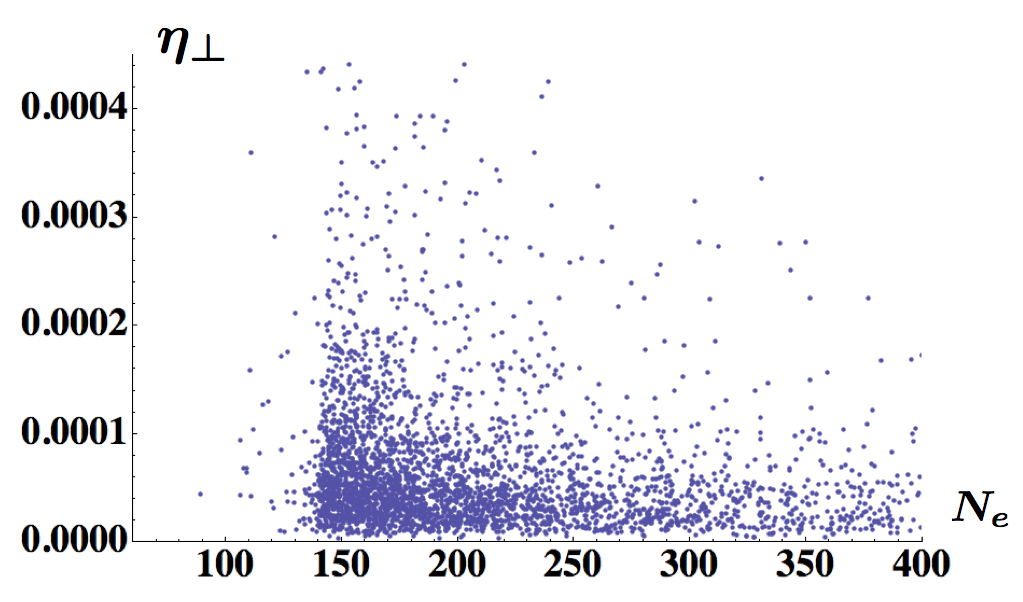} ~~~ &
\includegraphics[width=3.5in,angle=0]{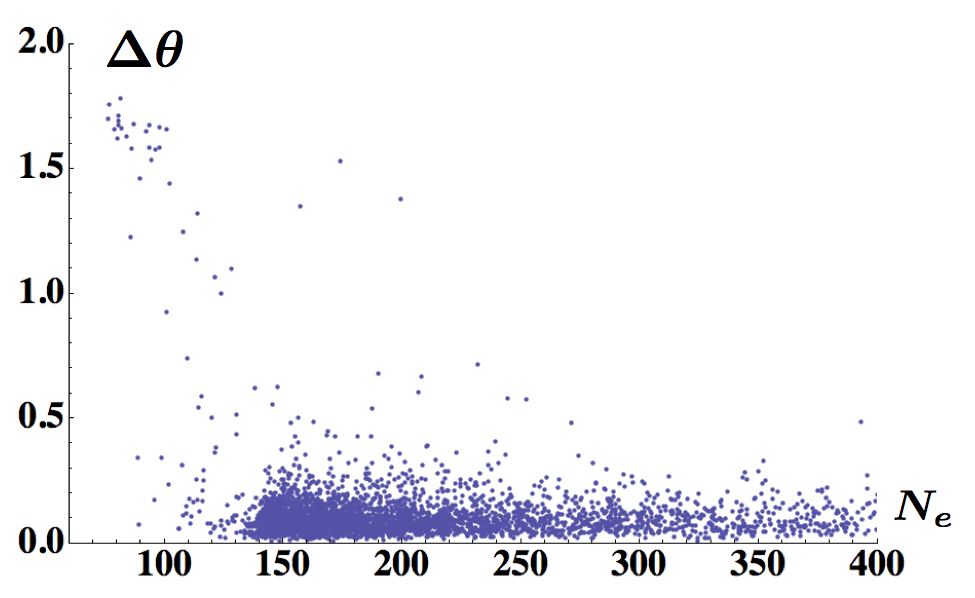}   \\
\end{array}$
\end{center}
\caption{Left panel: the slow roll parameter $\eta_{\perp}$ evaluated at Hubble crossing, versus the total number of e-folds, $N_e$,  for models with tilt $n_{s}$ consistent with observations. Right panel: the total turn of the trajectory, from six e-folds before Hubble crossing until the end of inflation, $\Delta \theta \equiv\int_{HC-6}^{end} dN_e\,\eta_{\perp}$, versus the total number of e-folds, $N_e$, for models with tilt $n_{s}$ consistent with observations.}
\label{turnarray5}
\end{figure}

It follows that for any realization consistent with observations, either (i) $N_e > 120$, and the tilt can be well-approximated by the single-field slow roll result, or (ii) the number of e-folds is constrained only by the solution of the horizon problem, and the tilt must differ substantially from the single-field slow roll result. Although case (ii) can occur, it is quite rare: multifield effects do tend to redden the spectrum, as noted
in \S\ref{multifieldtilt}, but given how blue the {\sf  naive} spectrum is, the multifield effect is typically not large enough to bring the {\sf exact} spectrum into agreement with observations (corresponding to points falling inside the orange band in Fig.~\ref{dnsmultifieldfigure1}).
In case (i), which describes 98.7\% of realizations consistent with observations, the 60 or more e-folds of inflation that precede the exit of the observed CMB generally suffice to damp out all
transient effects from the approach to the inflection point (this is reflected in Fig.~\ref{turnarray5}). Thus, although turning trajectories that imprint multifield effects in the curvature perturbations are present in a reasonable fraction (roughly 30\%) of realizations of inflation, they only represent 1.3\% of models with an observationally allowed tilt, in which these turns are generally complete long before the CMB exits the Hubble radius.

For the same reason, violations of slow roll are rare in realizations consistent with observations:
99\% of models with an allowable tilt have $-0.1< m_{\sigma}^2/H^2 < -0.01$.
Similarly, trajectories with substantial bending are uncommon once constraints on the tilt are imposed.
In the left panel of Fig.~\ref{turnarray5} we show $\eta_{\perp}$ at Hubble crossing for the observationally allowed ensemble, which is to be contrasted with the left panel of Fig.~\ref{turnarray}.  (In order to show the structure at small $\eta_{\perp}$, we have omitted a small number of points with  $\eta_{\perp} \sim 0.1$.)
The right panel of Fig.~\ref{turnarray5} shows the total turn of the trajectory,
to be contrasted with the right panel of Fig.~\ref{turnarray}.
Note that a handful of realizations allowed by constraints on the tilt do have significant bending.

\begin{figure}[!h]
  \begin{center}
    \includegraphics[width=5.5in,angle=0]{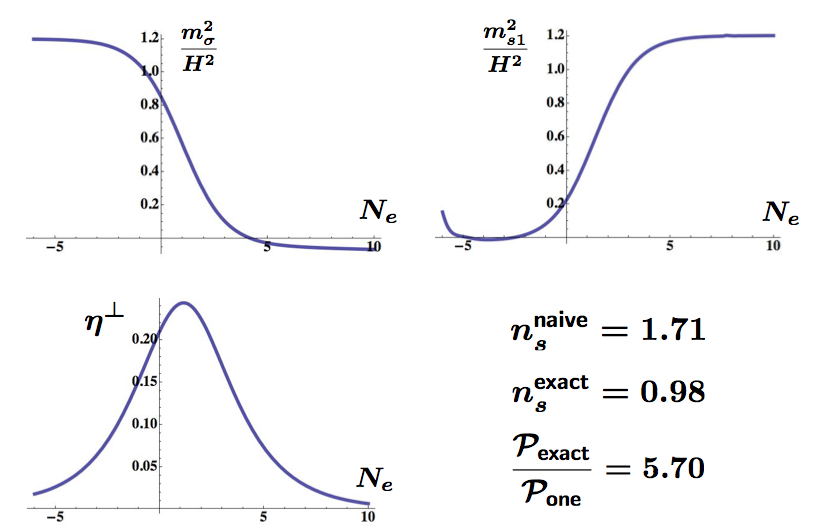}
    \caption{An example in which multifield effects --- specifically, {\sf two-field} effects --- are large, and the power spectrum is consistent with observations.
    This realization also appears in the lower right panel of Fig.~\ref{evolvearray}.}
    \label{Exampletrophy}
  \end{center}
\end{figure}

It is interesting to understand the characteristics of the realizations that have important multifield effects and are also consistent with constraints on the tilt.
Perhaps surprisingly, these realizations do not display extraordinary background behavior.
Of course, all such models have non-negligible bending of the trajectory around or after Hubble crossing, but the maximum value of $\eta_\perp$ along the inflationary evolution need not be greater than ${\cal O}(0.1)$ to generate significant multifield effects.  In fact, a majority display only a moderate degree of bending, $\eta_\perp \sim {\cal O}(0.05)$, albeit sustained over ten or more e-folds.  One distinctive difference compared to the general population is a substantially smaller value of the first entropic mass, $m_{s_1}^2$, at the time of Hubble crossing.  Whereas $m_{s_1}^2$ ranges from $0$ to $10$ in the full ensemble, typical values are of order $0.1$ in observationally allowed realizations with important multifield effects.  We show an example of such a realization in Fig.~\ref{Exampletrophy}.

\subsection{Non-Gaussianities}
\label{NGs}

Although so far we have only discussed the scalar power spectrum, a major motivation for the study of multifield models is the prospect of a detectable primordial bispectrum.
The D3-brane inflation scenario we have considered involves six fields that interact via Planck-suppressed couplings, leading to masses of order $H$ for each field.\footnote{Recall from \S\ref{spectrumobservations} that this is only a parametric estimate, and there is considerable additional structure in the scalar mass spectrum.}  Our setting therefore corresponds to a microphysical realization of {\it{quasi-single-field inflation}}, introduced by Chen and Wang in \cite{QSF}.  Quasi-single-field inflation occupies the middle ground between single-field inflation, in which any scalar fields other than the inflaton have masses $m \gg H$, and the simplest models of multifield inflation, in which all the fluctuating fields have masses $m \ll H$.

A striking feature of quasi-single-field inflation is the possibility of large non-Gaussianity, given apparently modest cubic couplings in the Lagrangian, and modest rates of turning, as measured by $\eta_{\perp}$.
Specifically, there is a contribution to the bispectrum scaling as \cite{QSF}
\begin{equation} \label{qsfestimate}
f_{NL} \sim \frac{1}{\sqrt{{\cal{P}}_{{\cal{R}}}}}\frac{V_{s_1 s_1 s_1}}{H}\eta_{\perp}^{3}\,,
\end{equation} where  $V_{s_1 s_1 s_1}$ denotes the third derivative of the potential with respect to the instantaneous first entropic direction.
Evidently, if $V_{s_1 s_1 s_1}$ is not very small compared to $H$ then $f_{NL}$ can be large, even for perturbatively small $\eta_{\perp}$.

We should point out that several important simplifying assumptions made in \cite{QSF} are not necessarily applicable in our setting.
In particular, $\eta_\perp$ and the first entropic mass squared, $m_{\s}^2$, were assumed to be constant during inflation.  Moreover, the authors of \cite{QSF} considered
a regime in which the multifield effects give subdominant contributions to the amplitude of the power spectrum 
(see also \cite{Pi:2012gf}).
Finally, it was assumed in \cite{QSF} that $\eta_\perp$ was small enough to be used as a perturbative expansion parameter.
This last point implies that the operator identified as leading in \cite{QSF} may not give the dominant contribution to $f_{NL}$ in our context. In the following, we will obtain a rough picture of the prospects for non-Gaussianity in our ensemble using the estimate (\ref{qsfestimate}).
Going beyond the approximations above is an interesting problem for the future.

Although quasi-single-field inflation is a plausible proposal at the level of field theory, it is reasonable to ask whether the particular assumptions required to achieve detectable non-Gaussianity are natural in the effective theories arising from well-motivated microscopic constructions.  For clarity, we distinguish two properties of the effective theory: first, that typical scalars --- including the entropic fluctuations --- have masses of order $H$; and second, that the cubic couplings $V_{s_1 s_1 s_1}$ are not too small\footnote{For $V_{s_1 s_1 s_1} \gtrsim H$, there is a risk of large radiative corrections to the masses.  We thank Daniel Baumann and Daniel Green for instructive discussions of this point.} compared to $H$.  There is very ample and long-established evidence for the first property in theories with spontaneously broken supersymmetry, as recently explained in detail in \cite{BaumannGreen}, and as noted above this is borne out in detail in our ensemble.  However, the scale of the cubic couplings is more subtle, as we now explain.

Consider a D3-brane in a conifold region of a de Sitter compactification of string theory.
Using a spurion superfield $X$ to represent the source of supersymmetry breaking, so that $|F_X|^2 = 3H^2 M_p^2$, the scalar potential for the D3-brane includes contributions of the form  \cite{Baumann10}
\begin{equation}
V =  \sum_{\Delta} c_{\Delta} \int d^4\theta X^{\dagger}X \left(\frac{\phi}{M_p}\right)^{\Delta}\, f_{\Delta}(\Psi)  =  \sum_{\Delta} 3 c_{\Delta} H^2 M_p^2 \left(\frac{\phi}{M_p}\right)^{\Delta}\,f_{\Delta}(\Psi)\,,
\end{equation} where $\Delta$ is an operator dimension, $c_{\Delta}$ is a Wilson coefficient that is expected to be of order unity, and $\phi = \sqrt{T_3} r$, with $T_3$ the D3-brane tension, is the canonically normalized field representing radial motion.  Here $f_{\Delta}(\Psi)$ is a (known) function of the dimensionless angular coordinates $\theta_1,\phi_1,\theta_2,\phi_2,\psi$, collectively denoted $\Psi$.

It will be critical to relate the  $\Psi$ to the canonically normalized angular fields, which we denote by $\varphi_{a}$, $a=1\ldots5$.
The metric on the angular manifold $T^{1,1}$ is not diagonal in the $\theta_1,\phi_1,\theta_2,\phi_2,\psi$ basis, but one can perform a coordinate transformation to obtain independent dimensionless fields, which we may write as $\hat{\Psi}_a$, $a=1\ldots5$.  The kinetic term for the $\hat{\Psi}_a$ takes the form ${\cal L} \supset  \sum_{a=1}^5 \phi^2 (\partial\hat{\Psi}_a)^2\,,$
so at a given radial location $\phi=\phi_{\star}$, the canonically normalized angular fields are given by $\varphi_{a} = \phi_{\star}\hat{\Psi}_a$.

The cubic coupling for three angular fields therefore takes the schematic form
\begin{equation}
V_{\varphi_{a} \varphi_{b} \varphi_{c}} \sim \sum_{\Delta} c_{\Delta} \frac{H^2}{M_p} \left(\frac{\phi_{\star}}{M_p}\right)^{\Delta-3}\,f^{abc}_{\Delta}(\Psi)\,,
\end{equation} where $f^{abc}_{\Delta}$ denotes the corresponding derivative of $f$, which is generally of order unity.
We observe that differentiating with respect to canonically normalized angular fields has introduced the factor $(M_p/\phi_{\star})^3$.
As many operators with $\Delta<3$ are present in the D3-brane Lagrangian, and the inflationary inflection point occurs at $\phi_{\star} \ll M_p$, the cubic couplings of the angular fields are therefore parametrically large compared to $\frac{H^2}{M_p}$.  This is worth emphasizing: if all fields had entered the Lagrangian on the same footing, and the scalar potential took the schematic form\footnote{This estimate relies on the assumption that the fields are moduli whose masses and cubic couplings vanish before supersymmetry breaking.  For D3-branes in the conifold, and more generally for moduli in string compactifications, this assumption is appropriate, but in more general settings larger cubic couplings need not be unnatural.  We thank Daniel Baumann for helpful comments on this point.}
\begin{equation}\label{polynomial}
V = H^2 M_p^2 P(\phi_1/M_p,\ldots \phi_N/M_p)\,,
\end{equation} with $P$ a general polynomial of $N$ fields $\phi_1/M_p, \ldots, \phi_N/M_p$, with Taylor coefficients of order unity, then the cubic couplings would scale as $V^{\prime\prime\prime} \sim H^2/M_p$.
In the conifold (or any other Calabi-Yau cone) we find a potentially dramatic enhancement compared to this `democratic' estimate.

The discussion so far has only addressed the couplings of the angular fields.  The cubic couplings of the radial field $\phi$ include terms of the form
\begin{equation} \label{cubicestimate}
V_{\phi\phi\phi} \sim  c_{\Delta}\Delta (\Delta-1)(\Delta-2) \frac{H^2}{M_p} \left(\frac{\phi_{\star}}{M_p}\right)^{\Delta-3}\,.
\end{equation}
Thus, if the $\Delta$ were all integers, we would have  $V_{\phi\phi\phi}  \lesssim H^2/M_p$, which is too small to contribute to detectable non-Gaussianity (see (\ref{qsfestimate2}) below). However, the operator dimensions are not all integers in the field theory dual to the conifold.  Inflection point inflation in this model typically results from a cancellation between a term with $\Delta=3/2$ and the well-known conformal coupling term with $\Delta=2$ (see \cite{Baumann08} for a discussion), which can occur at small values of $\phi$.
Differentiation of the term with $\Delta=3/2$ gives rise to a cubic coupling for $\phi$ that is of the same order as the cubic couplings of the angular fields.

To summarize, in a natural effective theory describing moduli that obtain their potential from gravitational-strength coupling to a source of supersymmetry breaking,
in which the scalar potential involves a generic polynomial of the form (\ref{polynomial}), involving only integer powers of the fields, with all fields entering on the same footing, the cubic couplings scale as $V^{\prime\prime\prime} \sim H^2/M_p$, and are far too small to generate detectable non-Gaussianity.  We have seen that D-brane inflation in the conifold necessarily modifies this simple picture, in two ways: the operator dimensions are not all integers, and the radial and angular fields play different roles in the potential.  The result is that the cubic couplings of all six fields are parametrically of order
\begin{equation}
V^{\prime\prime\prime} \sim \sum_{\Delta} \frac{H^2}{M_p} \left(\frac{\phi_{\star}}{M_p}\right)^{\Delta-3}\,.
\end{equation}

Having addressed the parametric scalings of the cubic couplings, we turn to examining the range of numerical values that the quantities appearing in equation (\ref{qsfestimate}) attain in our ensemble.  As a measure of the scale $H$, we use $\epsilon$: equation (\ref{qsfestimate}) can then be rewritten as
\begin{equation} \label{qsfestimate2}
f_{NL} \sim 10 \sqrt{\epsilon}\, \frac{V_{s_1 s_1 s_1} M_p}{H^2}\,\eta_{\perp}^{3} \,.
\end{equation}
The typical value is $\epsilon \sim 10^{-12}$, and there is a tail toward somewhat larger values: 7\% of realizations have $\epsilon > 10^{-8}$, though we found no examples with $\epsilon > 10^{-6}$.  The typical size of $\eta_{\perp}$ is $\sim 10^{-3}$, while $\eta_{\perp} <1$ in 99\% of realizations, and we found no example with $\eta_{\perp} > 3$.\footnote{These statements refer to the value of $\eta_\perp$ averaged from one e-fold before Hubble exit until one e-fold after Hubble exit, which can be considerably smaller than the maximum value of $\eta_\perp$ during this period.} The typical scale of the cubic coupling of the first entropic mode in our ensemble is $V_{s_1 s_1 s_1} \sim 10^3 H^2/M_p$, and as a conservative estimate we take $V_{s_1 s_1 s_1} \lesssim 10^4 H^2/M_p$.  Assembling these results, we find that $f_{NL} >10$ in less than $1\% \times 7\% = .07\%$ of all realizations.

The results above apply to the full ensemble, but non-Gaussianity is far less likely in the ensemble consistent with constraints on the tilt: 97 \% of observationally allowed models have $\eta_{\perp} \lesssim 10^{-3}$,
and  99\% have $\epsilon < 10^{-13}$, so that there is a negligible prospect of detectable non-Gaussianity in such models.

It is straightforward to extend our analysis to the quartic couplings, and correspondingly to estimate the overall amplitude $t_{NL}$ of the trispectrum.  We rewrite the result \cite{QSF}
\begin{equation} \label{tnlestimate}
t_{NL} \sim {\rm max} \left( \frac{1}{{\cal{P}}_{{\cal{R}}}} \left( \frac{V_{s_1 s_1 s_1}}{H} \right)^2\eta_{\perp}^{4} , \frac{1}{{\cal{P}}_{{\cal{R}}}} V_{s_1 s_1s_1s_1}   \eta_{\perp}^{4}   \right)  \,,
\end{equation}
where $V_{s_1 s_1 s_1 s_1}$ denotes the fourth derivative of the potential with respect to the instantaneous first entropic direction, as
\begin{equation} \label{tnlestimate2}
t_{NL} \sim 100 \, \epsilon\, \eta_\perp^4\, {\rm max} \left( \left( \frac{V_{s_1s_1s_1}M_p}{H^2} \right)^2  ,\frac{ V_{s_1s_1s_1s_1} M_p^2}{H^2}   \right)  \,.
\end{equation}
We find that typically $V_{s_1 s_1 s_1 s_1} \lesssim 10^8 H^2/M_p^2$, from which we deduce that there is a very limited possibility of a detectable trispectrum ($t_{NL} \gtrsim 1000$) in the full ensemble, but the corresponding probability is negligibly small in the ensemble consistent with constraints on the tilt.

In conclusion, our results suggest that significant non-Gaussianity ($f_{NL} \gtrsim 10$ and/or $t_{NL} \gtrsim 1000$) on scales far outside the present Hubble radius is possible, albeit somewhat rare, in warped D-brane inflation; but  correspondingly large non-Gaussianity is very rare on observable angular scales.
Even so, a dedicated study of non-Gaussianity in D-brane inflation --- or in some other well-motivated microphysical realization of quasi-single-field inflation --- would be worthwhile.

\section{Conclusions} \label{conclusions}

We have determined the scalar power spectrum that results when inflation takes place in a random six-field potential.
The potential in question governs the motion of a D3-brane in a conifold region of a string compactification, and was derived and extensively studied in \cite{Baumann10}.
The signal property for this work is not the string theory  provenance of the inflaton action, but merely the fact that the action is natural: the terms in the potential correspond to Planck-suppressed operators with coefficients of order unity, and the masses are of order $H$, as expected for moduli with  gravitational-strength couplings to a source of supersymmetry breaking. For this reason, we believe our analysis is representative of a general class of multifield potentials that are natural in the Wilsonian sense, even though aspects of the conifold geometry do influence our results.

The essence of our approach to computing the primordial perturbations was the comparison between numerical integration of the exact equations for the linearized perturbations, making no approximation, and an array of truncated models omitting one or more of the entropic modes.  This allowed us to determine the extent to which various approximations and truncations --- including the slow roll approximation, the single-field truncation, and the two-field truncation --- capture the physics of a generic realization of inflation.  As we made no slow roll or slow turn approximation, we were able to characterize phenomena that are common in our ensemble, but more rarely seen in analytic treatments.

We began by determining the spectrum of scalar masses using a random matrix model, building on \cite{MMW}, and demonstrated excellent agreement with simulations of the full potential.  We found that at the time of Hubble exit, one entropic mode is typically light enough to fluctuate, but before the end of inflation an adiabatic limit is reached, and one can predict the late-time curvature perturbation without modeling the details of reheating.

Our results for the perturbations are usefully divided into characterizations of the ensemble of realizations of inflation that give rise to at least 60 e-folds of inflation, but are otherwise unconstrained, and characterizations of the ensemble of realizations of inflation that give rise to at least 60 e-folds of inflation, and are also consistent with observations of the CMB temperature anisotropies: conditioning on the approximate scale-invariance required by observations drastically changes the outcome.

For the ensemble of inflationary models that solve the horizon problem but are not required to give nearly scale-invariant density perturbations, multifield effects are often significant: multifield corrections to the spectrum are at least at the 1\% level in roughly 30\% of realizations, and exceed 100\%
in 10\% of realizations.  We also found that {\it many-field} contributions to the perturbations --- by which we mean effects that cannot be described by the two-field truncation, which retains only the instantaneous adiabatic and first entropic modes --- are similarly common: many-field corrections exceed 1\% in 18\% of realizations, and exceed 100\% in 6\% of realizations.
Most models with substantial multifield effects have either large two-field effects or large many-field effects, but not both.
Finally, we observed that
the exact scalar power can be smaller than the power calculated in the single-field truncation: this is a consequence of trajectories that turn quickly at the time of Hubble crossing.

Turning now to the restricted ensemble of realizations that are consistent with the WMAP7 constraints on the tilt
of the scalar power spectrum, corresponding to 21\%
of the total ensemble, we find a qualitatively different picture.  When inflation occurs near an approximate inflection point, as it does in the class of potentials studied here, the tilt depends on whether the observed CMB exits the Hubble radius before or after the inflaton descends past the inflection point.  When the 60 e-fold mark occurs above the inflection point, the curvature of the potential is positive and the spectrum computed in the single-field slow roll approximation is blue, which is inconsistent with observations.  Therefore, the allowable potentials are those in which 60 or more e-folds occur below the inflection point.  In such potentials, there is a prolonged phase of inflation before the observed CMB exits the Hubble radius: namely, the inflation occurring above the inflection point.  This phase generally suffices to damp out all transients from the onset of inflation, {{\it including the curving trajectories that can give rise to multifield effects in the perturbations.}}  For this simple reason, although multifield contributions to the perturbations are commonplace in random inflection point models, for more than 98\% of models with an observationally allowed tilt these contributions arise on angular scales that are far outside the present Hubble radius.  We conclude that for inflection point models of the sort studied here, multifield effects in the observable perturbations are uncommon in realizations consistent with present limits on scale-invariance.  We stress that the multifield effects themselves are {\it{not}} responsible for the problematic deviations from scale invariance: in fact, the spectrum calculated in the single-field approximation is blue, and multifield effects shift it toward scale invariance.  Instead, multifield effects generally arise from transients from the onset of inflation, and at the onset of inflation at an inflection point, the dominant single-field component of the spectrum is often, but not always, unacceptably blue.

The ensemble we have studied provides a concrete microphysical realization of quasi-single-field inflation.  The couplings in the inflaton action arise from Planck-suppressed operators, and the cubic and quartic interactions of the entropic modes, although enhanced by contributions from an operator with dimension $\Delta=3/2$, as well as by the intrinsic asymmetry between the radial and angular directions of the conifold, are only occasionally sufficiently large to produce detectable non-Gaussianity along the lines envisioned in \cite{QSF}.
More generally, the suppression of transients resulting from constraints on the tilt ensures, as explained above, that in a majority of realizations consistent with observations of the spectrum,  the single-field slow roll approximation is valid.  Of course, the perturbations are Gaussian to good approximation in all such cases.  Understanding the generality of this finding in well-motivated effective theories is an important problem for the future.

We expect that our methods, and some aspects of our findings, have   broad applicability.
Our present understanding of string compactifications suggests considering ensembles of effective theories with numerous moduli and spontaneously broken supersymmetry.  In this work we have characterized the cosmological signatures of one such ensemble.  We have seen that when inflation occurs at an inflection point in a many-field potential, then even though it is common for multiple fields to fluctuate, and ultimately to contribute to the curvature perturbations, these contributions are most often on unobservably large angular scales.  The signatures of the model are therefore often indistinguishable from those of a single-field inflection point scenario.  Nevertheless, it remains to be seen which other universality classes of multifield inflation may exist, and it is reasonable to anticipate very different conclusions when the number of fields is large.

\subsection*{Acknowledgements}
We are particularly indebted to Daniel Baumann and David Seery for insightful comments on a draft of this work.
We thank Niayesh Afshordi, Thomas Bachlechner, Rachel Bean, Xingang Chen, Sera Cremonini, Mafalda Dias, Richard Easther, Jonathan Frazer, Ghazal Geshnizjani, Daniel Green, Kazuya Koyama, David Langlois, Louis Leblond, David Marsh, Enrico Pajer, Krzysztof Turzy\'nski, Yi Wang, Scott Watson, Dan Wohns, and Timm Wrase for helpful discussions of related topics.
L.M.\ and S.R-P.\ thank the Isaac Newton Institute and the organizers of String Phenomenology 2012 for providing a stimulating setting for the completion of this work.
The research of L.M.\ was supported by the Alfred P. Sloan Foundation, by an NSF CAREER Award, and by the NSF under grant PHY-0757868. S.R-P.\ is supported by the STFC grant ST/F002998/1 and by the Centre for Theoretical Cosmology.  Some of our simulations were performed on the COSMOS supercomputer, which is funded by STFC, BIS and SGI.


\end{document}